\documentclass[12pt]{elsarticle}
\graphicspath{{Figures/}}
\usepackage{amsmath,amssymb,amsthm,amsfonts}
\usepackage{caption}
\usepackage[title]{appendix}
\usepackage{xcolor}
\setcitestyle{authoryear,open={(},close={)}}

\usepackage{graphicx,epsfig}
\usepackage{psfrag}
\usepackage{xcolor}
\usepackage[hmargin=1in,vmargin=1in]{geometry}
\usepackage{stabular,morefloats}
\usepackage{textcomp}
\usepackage{titlesec}
\usepackage{colortbl}
\usepackage{xcolor}
\usepackage{multirow}
\usepackage{mathrsfs}
\usepackage{tabularx}
\usepackage{comment}
\usepackage{hyperref}
\usepackage{multirow,tabularx,afterpage,fancyhdr}
\usepackage{placeins}
\titleformat{\subsubsection}
  {\normalfont\selectfont}{\thesubsubsection}{1em}{}
\numberwithin{equation}{section}
\usepackage{float}
\usepackage{xcolor}
\usepackage{soul}
\definecolor{lightblue}{rgb}{0,0,0.1}
\sethlcolor{lightblue}

\usepackage[subrefformat=parens,labelformat=parens]{subfig}

\colorlet{Changes@Color}{black}
\makeatletter
\renewcommand{\@biblabel}[1]{[#1]\hfill}
\makeatother

\makeatletter
\def\ps@pprintTitle{%
 \let\@oddhead\@empty
 \let\@evenhead\@empty
 \def\@oddfoot{}%
 \let\@evenfoot\@oddfoot}
\makeatother

\begin{document}

\title{A physics-informed neural network for modeling fracture without gradient damage: formulation, application, and assessment}

\author[main]{Aditya Konale}
\author[main,main2]{Vikas Srivastava\corref{cor1}}
\ead{vikas\_srivastava@brown.edu}
\cortext[cor1]{Corresponding author}
\address[main]{School of Engineering, Brown University, Providence, RI 02912, USA}
\address[main2]{Institute for Biology, Engineering and Medicine, Brown University, Providence, RI 02912, USA}

\begin{abstract}
Accurate computational modeling of damage and fracture remains a central challenge in solid mechanics. The finite element method (FEM) is widely used for numerical modeling of fracture problems; however, classical damage models without gradient regularization yield mesh-dependent and usually inaccurate predictions. The use of gradient damage with FEM improves numerical robustness but introduces significant mathematical and numerical implementation complexities. Physics-informed neural networks (PINNs) can encode the governing partial differential equations, boundary conditions, and constitutive models into the loss functions, offering a new method for fracture modeling. Prior applications of PINNs have been limited to small-strain problems and have incorporated gradient damage formulation without a critical evaluation of its necessity. Since PINNs in their basic form are meshless, this work presents a PINN framework for modeling fracture in elastomers undergoing large deformation without the gradient damage formulation. The PINN implementation here does not require training data and utilizes the collocation method to formulate physics-informed loss functions. We have validated the PINN’s predictions for various defect configurations using benchmark solutions obtained from FEM with gradient damage formulation. The crack paths obtained using the PINN are approximately insensitive to the collocation point distribution. This study offers new insights into the feasibility of using PINNs without gradient damage and suggests a simplified and efficient computational modeling strategy for fracture problems. The performance of the PINN has been evaluated through systematic variations in key neural network parameters to provide an assessment and guidance for future applications. The results indicate that in elastomer-like materials, fracture progression and path evolution are primarily governed by the damage initiation criterion. This study provides motivation for extending PINN-based approaches to a broader class of materials and damage models in mechanics.

\end{abstract}

\vspace{0.25in}


\maketitle

\section{Introduction}\label{sec1}
The ability to computationally model and predict damage and fracture of materials and structures is essential for enhancing scientific understanding of failures and enabling robust real-world engineering solutions. The finite element method (FEM) is extensively used for numerical modeling of fracture for a variety of problems \citep{FEMfracture1, FEMfracture2, FEMfracture3}. A straightforward way to model damage and fracture is by using a local damage field variable in FEM without incorporating non-local gradient damage terms. In FEM, the stiffness of an element is reduced to zero when the damage field variable at an integration point reaches a critical value (fully damaged value at the local point). Solutions from FEM without gradient damage have been shown to exhibit significant mesh-dependency. The predicted crack propagation path can be nonphysical and inaccurate if the finite element mesh is not aligned with the actual path, which can be difficult to guess a priori \citep{Elementdellim1, Elementdel1}. The nonlocal gradient damage term in mesh-based numerical modeling is needed to regularize for mesh-independent solutions. 
Computational models and methods that can predict crack nucleation, propagation, and coalescence in arbitrary, complex geometries without requiring the mesh to be aligned with a priori unknown crack paths are necessary for solving a general class of problems. For numerical reasons, FEM with gradient damage considers a ``smeared crack approach", in which the mathematical framework for damage depends on a damage field variable $d$ along with its gradient $\nabla d$. A length scale parameter $l$ is also used over which $d$ varies between zero and one \citep{Miehefirstpaper}. FEM with gradient damage can provide mesh-independent results, provided that the typical element size is less than 0.2$l$ \citep{Anandquasibrittle}. Numerous works in the literature have used FEM with gradient damage to model damage initiation, growth, and complete failure in a broad range of materials \citep{Anand(2021)flowrule, KUHL2000, Shabnamphasefield, Guocytoskeletalpf, Dolbowpf2, KUMAR1, KONALEMoM}. However, the mathematical and numerical implementation complexity of FEM with gradient damage is significantly higher compared to that of the simple FEM without gradient damage. An additional degree of freedom $d$ has to be considered at each point in FEM with gradient damage, and the corresponding partial differential equation for damage evolution has to be solved. This involves spatial gradient calculations and the development of complex user-defined elements and user subroutines for implementation within FEM programs \citep{Anandquasibrittle, Shabnamphasefield}.

Neural networks have been increasingly explored as computational methods to solve engineering mechanics problems previously difficult to analyze \citep{Niu2022UltrasoundNetwork, Niu2022SimulationMeasurements}. Physics-informed neural networks (PINNs) have emerged in the recent literature as a numerical method for solving mechanics problems \citep{PINNsoldmech1, PINNsoldmech2, PINNsoldmech3, PINNsoldmech4, SijunplasticityPINN, PINNfracture8, PINNhyperelasticity1, PINNhyperelasticity2, PINNhyperelasticity4, PINNhyperelasticity5, Bai2025, DONG2025, GUO2025}. Although PINNs are relatively new compared to FEM \citep{PINNvsFEM1}, they have certain advantages, which include their open-source nature and relative ease of implementation \citep{PINNvsFEM2}. PINNs can solve partial differential equations (PDEs) by
encoding the mathematical information explicitly into the loss functions without a need for training data. The application of PINNs for mechanics and their advantages and disadvantages with respect to the conventional numerical methods (e.g., FEM) are summarized in Table \ref{pinnsformechanicstablefig}.

Some recent works have focused on the application of PINNs to damage and fracture problems in the small-strain regime \citep{PINNfracture1, PINNfracture2, PINNfracture3, PINNfracture4, PINNfracture5, PINNfracture7, PINNfracture8, PINNfracture9}. These works directly adopt the relatively complex mathematical formulation of the gradient damage method to develop the physics-based loss functions. Loss functions in the form of variational energy are more common than separate loss functions that individually enforce the underlying partial differential equations and boundary conditions. Variational energy-based loss functions require numerical integration over the domain to evaluate energy integrals. Variational energy-based PINNs have been developed in \citep{PINNfracture2, PINNfracture3, PINNfracture5}, which solve the problem incrementally using transfer learning and without training data. Goswami and co-workers \citep{PINNfracture1} proposed a variational form of DeepONet (V-DeepONet), which is trained by imposing the governing equations in a variational form with labeled data. The V-DeepONet, once trained, can rapidly predict the global solution for any initial crack configuration and loading on that domain. Improvements in the V-DeepONet have been presented in \citep{PINNfracture9}. Yi and co-workers proposed a mechanics-informed, model-free symbolic regression framework for solving small-strain fracture problems. 

The PINN and physics-informed machine learning-based damage and fracture modeling work in the literature is limited to small strains, and these PINNs directly incorporate the gradient damage formulation. Further, due to the meshless nature of PINNs \citep{PINNmeshless1}, the gradient damage formulation utilized primarily for numerical reasons (to eliminate mesh dependency in FEM) is not necessary.

\begin{table*}[h!]
    \centering
    \begin{tabular}{|m{35em}|}
        \hline
        \hspace{5 cm}\textbf{PINNs for Mechanics}\\ \vspace{2 mm}
        •	Spatial coordinates, time, PDE parameters, etc., as inputs to the NN.\\
        •	Displacements, stress tensor components, field variables, PDE parameters, etc., as the NN outputs.\\ 
        •	Loss function divided into two parts: data loss (IC/BCs, and any experimental data) and residual/governing equation loss (computed using the strong or weak formulation of PDEs).\\
        •	Loss function minimization (e.g., MSE, MAE, Sobolev norm) for obtaining the optimal parameters of the NN. The loss is first computed point-wise (for strong formulation) over identical and independently distributed residual (collocation) points as well as IC/BC points.\\
        \hline
    \end{tabular}
\end{table*}

\begin{table}[H]
    \centering
    \begin{tabular}{|m{17.2em}|m{17.2em}|}
        \hline
        \hspace{2.5 cm}\textbf{Advantages} & \hspace{2.3 cm}\textbf{Disadvantages}\\
        \hline
        Relative ease of implementation from scratch and open-source nature compared to the conventional numerical methods. & Accuracy of the solution depends on various hyperparameters of the neural network. Lack of guidance for a general class of mechanics problems. Rigorous approaches such as grid search can be used for parameter tuning. \\
        \hline
        Mesh-free nature and automatic differentiation (AD) capabilities. AD computes exact gradients as the activation functions have analytical gradient forms. & Forward problems without training data need significantly larger computation time than conventional numerical methods. \\
        \hline
        Non-linear activation functions (e.g., tanh with guaranteed differentiability) are more expressive than conventional method basis functions. & Non-convex, non-linear, and multi-objective nature of loss functions often results in slow convergence.\\
        \hline
        Combination with data-driven learning is promising. Training can be achieved for sparse and high-dimensional data. & “Black box” nature lacks interpretability but can be used as a surrogate model for “Gray box” learning, such as SR/SINDy.\\
        \hline
        Handle ill-posed inverse problems better than conventional methods due to the incorporation of physics as a regularizer - mimetic to Tikhonov regularization. & Application to multi-scale and multi-physics problems is challenging.\\
        \hline
        Can recover missing physics or constitutive relations. Unlike conventional methods, IC/BCs can be partially described. & Detailed validations for a broad class of problems are still in their infancy. \\
        \hline
    \end{tabular}
    \captionsetup{labelfont={small}} 
    \caption{\small \textbf{A summary of PINNs for mechanics and their advantages and disadvantages with respect to the conventional numerical methods like Finite Element Method.} \\
    (IC - initial conditions, BC - boundary conditions, MSE - mean squared error, MAE - mean absolute error, SR - symbolic regression, SINDy - sparse identification of nonlinear dynamics)}
    \label{pinnsformechanicstablefig}
\end{table}

The main contributions of this work are: 

\begin{enumerate}[(i)]
\item We have proposed a PINN without gradient damage for modeling fracture. Specifically, it does not require training data and does not have the gradient damage method-associated mathematical and numerical implementation complexities. 

\item The PINN is formulated to model large deformation fracture of elastomers. To the best of our knowledge, this has not been attempted in the literature. 

\item We have validated the predictive capabilities of the PINN (PINN without gradient damage) for various defect configurations using baseline solutions obtained from FEM with gradient damage. The proposed modeling framework and neural network architecture should motivate PINNs without gradient damage to model damage and fracture in a wide range of materials. 

\item The PINN as a fracture modeling method is evaluated through systematic variations in key parameters. The PINN crack path predictions are approximately insensitive to the distribution of collocation points. We also show that relatively computationally inexpensive one-time step PINN solutions provide reasonably accurate crack paths for rate-independent materials under monotonic loading.

\end{enumerate}

Recently, large deformation fracture of elastomers was accurately modeled in \citep{KONALEMoM} using the gradient damage framework with FEM \citep{KONALEMoM}. We use the critical free energy density-based damage initiation criteria proposed in \citep{KONALEMoM} for our PINN model. In the presence of multiple defects, crack path prediction is complex. Accurate crack path prediction is important for structural applications. We apply our PINN without gradient damage to predict the crack paths for various defect configurations.
 The plan of the rest of the paper is as follows. The constitutive equations for the modeling of large deformation fracture of elastomers without gradient damage are presented in Section 2, and the corresponding PINN formulation is presented in Section 3. The results of the PINN predictions are validated in Section 4 with a variety of fracture problems using various defect configurations. In Section 5, we provide an assessment of the PINN's performance for various neural network parameter choices and conclude the paper in Section 6. 

\section{Modeling framework and constitutive equations}

The proposed PINN is applied to elastomers in this work to demonstrate its predictive capabilities. We first summarize the basic large-deformation modeling framework and then discuss the damage modeling approach used in this work. The deformation gradient \textbf{F}, the right Cauchy-Green deformation tensor \textbf{C} and the left Cauchy-Green deformation tensor \textbf{B} are defined as 

\begin{equation}
    \textbf{F}(\textbf{X},t) = \dfrac{\partial \textbf{x}(\textbf{X},t)}{\partial \textbf{X}}, \quad \textbf{C} = \textbf{F}^{\text{T}} \textbf{F}, \quad \textbf{B} = \textbf{F} \textbf{F}^{\text{T}},
\label{F, C, B definition}    
\end{equation}

\hspace{-0.23 in}with \textbf{X} and \textbf{x} being the undeformed and deformed position vectors of a material point, respectively. $\textbf{u} = \textbf{x} - \textbf{X}$ is the displacement of a material point at $\textbf{X}$ and $t$ is the time instant under consideration. The specific form for the referential undamaged free energy density function $\psi$ considered here for the elastomer is compressible Neo-Hookean. 

\begin{equation}
    \psi = \dfrac{\mu}{2}(I_{1}-3) + \dfrac{K}{2}(J-1)^{2}, \quad I_{1} = \text{trace} (\textbf{C}_{dis}), \textbf{C}_{dis} = J^{-\frac{2}{3}} \textbf{C} \: \text{ and } \: J = \text{det} \textbf{F}, 
\label{NeoHookean function}    
\end{equation}

\hspace{-0.23 in}where $\mu$, $K$ are the ground state shear and bulk modulus, respectively. $\textbf{C}_{dis}$, $I_{1}$, $J$ are the distortional part of $\textbf{C}$, the first invariant of $\textbf{C}_{dis}$ and determinant of \textbf{F}, respectively. The approach discussed in this section should apply to any functional form (hyperelastic model) of choice for $\psi$. A damage variable $d(\textbf{X})$ is considered to model damage, which can take the values of either 0 (undamaged) or 1 (fully damaged). This is equivalent to instantaneous damage growth upon initiation at a material point experimentally observed in elastomers \citep{EPDMrubber, Parabolicshape2}. A material point is considered fully damaged ($d$ = 1) if $\psi^{+}$ at a material point at any deformation step exceeds a critical value $\psi^{+}_{cr}$, i.e., 

\begin{equation}
    d=\begin{cases}
\begin{aligned}
			&0 \hspace{0.2in} \hspace{0.1in} &&\text{if}  \hspace{0.1in} \psi^{+} \leq \psi^{+}_{cr},\\
			&1  &&\text{if} \hspace{0.1in} \psi^{+} > \psi^{+}_{cr}.
\end{aligned}
	\end{cases}
\label{damage criterion} 
\end{equation}

\hspace{-0.23 in}$\psi^{+}$ here is the part of $\psi$ which is distortional and tensile dilatational, i.e.,  

\begin{equation}
    \psi^{+}=\begin{cases}
\begin{aligned}
			&\dfrac{\mu}{2}(I_{1}-3) + \dfrac{K}{2}(J-1)^{2} \hspace{0.2in} \hspace{0.1in} &&\text{if}  \hspace{0.1in} J>1,\\
			&\dfrac{\mu}{2}(I_{1}-3)  &&\text{if} \hspace{0.1in} J\leq 1.
\end{aligned}
	\end{cases}
\label{psio+ definition} 
\end{equation}

\hspace{-0.23 in}This split accounts for damage growth being driven only by distortional and tensile dilatational deformation \citep{HDsplit1, HDsplit2}. To ensure irreversibility of damage, it is required that

\begin{equation}
    \text{if} \:\: d(\textbf{X},t) =1, \: \text{then} \:\: d(\textbf{X}, t+dt) = 1,
\label{damage irreversibility} 
\end{equation}

\hspace{-0.23 in}with $dt$ being an infinitesimal increment in $t$. The first Piola-Kirchhoff stress $\textbf{P}$ and Cauchy stress $\textbf{T}$ tensors are then given as

\begin{equation}
    \textbf{P} = J \textbf{T} \textbf{F}^{-\text{T}}, \:\:\:\: \textbf{T}=\begin{cases}
\begin{aligned}
			&(1-d)\Big[\dfrac{\mu}{J}(\textbf{B}_{dis})_{0} + K(J-1)\textbf{I}\Big] &&\text{if} \hspace{0.1in} J>1,\\
            &(1-d) \Big[\dfrac{\mu}{J}(\textbf{B}_{dis})_{0} \Big]+ K(J-1)\textbf{I} &&\text{if} \hspace{0.1in} J\leq1,
\end{aligned}
	\end{cases}
\label{Cauchy stress neohookean}
\end{equation}

\hspace{-0.23 in}with $(\textbf{B}_{dis})_{0}$ denoting the deviatoric part of $\textbf{B}_{dis}$.

\section{PINN for large deformation fracture modeling}
\label{PINN discussion}

\subsection{Neural network architecture}

We now present the procedure for constructing the PINN to evaluate fracture solutions in arbitrary geometries subjected to large deformations. The PINN has an input layer, one or multiple hidden layers, and an output layer. Two-dimensional problems are solved using the plane strain assumption. The problems are cast in a Lagrangian framework with $\textbf{X} = \{X_{1}, X_{2}\}$ denoting the coordinates of a material point in the reference configuration. A mixed formulation is used wherein the neural network approximates the first Piola-Kirchhoff stress field $\hat{\textbf{P}}(\textbf{X}, t)$ along with the displacement field $\hat{\textbf{u}}(\textbf{X}, t)$. The mixed formulation avoids the increased computational cost associated with higher-order partial derivative computations in the displacement-only formulation, along with accuracy and convergence issues \citep{SijunplasticityPINN}. In addition, the undamaged free energy density $\hat{\psi}(\textbf{X}, t)$, considered as a state variable, is also approximated by the PINN. All trainable PINN parameters, including the weights and biases in the neurons of the network, are denoted by the vector $\boldsymbol{\theta}$. The PINN formulation can then be written as

\begin{equation}
    (\hat{\textbf{u}}, \hat{\textbf{P}}, \hat{\psi}) = \mathcal{N}\mathcal{N}(\textbf{X}, t, \boldsymbol{\theta}),
\label{PINN basic formulation}
\end{equation}

\hspace{-0.23 in}with $\hat{}$ denoting direct outputs of the neural network. Specifically, $\hat{\textbf{u}} = \{u_{1}, u_{2}\}$, $\hat{\textbf{P}} = \{P_{11}, P_{12}, P_{21}, P_{22}\}$. Hence, there are a total of 7 outputs for a given $\textbf{X}$ and $t$. Damage $d(\textbf{X}, t)$ in this modeling framework can take values of either 0 or 1 as seen through equation \eqref{damage criterion}. We evaluate the approximate field $\tilde{d}$ during training using $\tilde{\psi}(\hat{u})$ [using equation \eqref{psio+ definition}] and equations \eqref{damage criterion}, \eqref{damage irreversibility} as

\begin{equation}
    \tilde{d}(\textbf{X},t)=\begin{cases}
\begin{aligned}
			&0 \hspace{0.2in} \hspace{0.1in} &&\text{if}  \hspace{0.1in} \tilde{\psi}^{+}(\textbf{X},t) \leq \psi^{+}_{cr},\\
			&1  &&\text{if} \hspace{0.1in} \tilde{\psi}^{+}(\textbf{X},t) > \psi^{+}_{cr},\\
            &1  &&\text{if} \hspace{0.1in} \tilde{d}(\textbf{X},t-dt) = 1.
\end{aligned}
	\end{cases}
\label{PINN damage criterion} 
\end{equation}

\subsubsection{Loss function construction}
\label{lossfunctionconstruction}

A general boundary value problem involving a continuum solid body can be mathematically described as 

\begin{equation}
\begin{split}
    &\underbrace{\text{Div}\textbf{P} + \textbf{b}_{R} = \rho_{R} \ddot{\textbf{x}}}_{\text{Balance of linear momentum}} \:\: \text{and} \:\: \underbrace{\textbf{P}\textbf{F}^{\text{T}} = \textbf{F}\textbf{P}^{\text{T}}}_{\text{Balance of angular momentum}}\: \text{in} \: \Omega, \\ & \:\:\:\:\:\:\:\:\:\:\:\:\:\:\:\:\:\:\:\:\:\:\:\:\:\:\:\:\:\:\:\:\:\:\:\:\:\:\:\:  \textbf{u}=\overline{\textbf{u}} \: \text{on} \: \Gamma_{u}, \\ & \:\:\:\:\:\:\:\:\:\:\:\:\:\:\:\:\:\:\:\:\:\:\:\:\:\:\:\:\:\:\:\:\:\:\:\:\textbf{P}\textbf{n}_{R} = \overline{\textbf{t}}_{R} \: \text{on} \: \Gamma_{t}.
\end{split}
\label{General BVP}
\end{equation}

\hspace{-0.23 in}$\Omega$ denotes the problem's referential domain. $\textbf{b}_{R}$, $\rho_{R}$ are the referential body force per unit volume and referential density, respectively. $\textbf{u}$ is the unknown displacement field and $\overline{\textbf{u}}$ is the prescribed displacement field on part of the boundary $\Gamma_{u}$. $\textbf{P}$ is the unknown first Piola-Kirchhoff stress field, $\textbf{n}_{R}$ is the outward unit normal vector to the referential configuration surface and $\overline{\textbf{t}}_{R}$ is the prescribed traction field on the part of the boundary $\Gamma_{t}$. The PINN consists of a physics-informed loss function encoding the discrete version of the underlying boundary value problem, in this case, equation \eqref{General BVP}. $\Omega$ is discretized using $N_{d}$ points located at $\textbf{X}^{i}_{d}, i=1, 2,....,N_{d}$. Similarly, $\Gamma_{u}$ and $\Gamma_{t}$ are discretized using $N_{u}$ points ($\textbf{X}^{i}_{u}, i=1, 2,....,N_{u}$) and $N_{t}$ points ($\textbf{X}^{i}_{t}, i=1, 2,....,N_{t}$), respectively. The time domain is discretized into $N_{t}$ time steps ($t^{j}, j=1,2,....,N_{t}+1$). Neglecting body forces, the balances of linear and angular momentum in equation \eqref{General BVP} can be collocated at a given time instant $t^{j}$ as

\begin{equation}
\begin{split}
&\:\:\:\:\:\:\:\:\:\:\:\:\:\:\:\:\:\:\:\:\:\:\:\:\:\:\:\:\:\:\text{Div}\hat{\textbf{P}}(\textbf{X}^{i}_{d}, t^{j}) = \rho_{R}\ddot{\hat{\textbf{u}}}(\textbf{X}^{i}_{d}, t^{j}), \\& \hat{\textbf{P}}(\textbf{X}^{i}_{d}, t^{j}) \Big(\textbf{I}+\dfrac{\partial \hat{\textbf{u}}(\textbf{X}^{i}_{d}, t^{j})}{\partial \textbf{X}}\Big)^{\text{T}} = \Big(\textbf{I}+\dfrac{\partial \hat{\textbf{u}}(\textbf{X}^{i}_{d}, t^{j})}{\partial \textbf{X}}\Big) \hat{\textbf{P}}^{\text{T}}(\textbf{X}^{i}_{d}, t^{j}).
\label{Balance equations collocation} 
\end{split}
\end{equation}

\hspace{-0.23 in}The displacement and traction boundary conditions at a time instant $t^{j}$ can be applied similarly by the following collocations

\begin{equation}
    \hat{\textbf{u}}(\textbf{X}^{i}_{u}, t^{j}) = \overline{\textbf{u}}(\textbf{X}^{i}_{u}, t^{j}),
\label{Displacement bc collocation}
\end{equation}

\begin{equation}
    \hat{\textbf{P}}(\textbf{X}^{i}_{t}, t^{j}) \textbf{n}_{R}(\textbf{X}^{i}_{t}) = \overline{\textbf{t}}_{R}(\textbf{X}^{i}_{t}, t^{j}).
\label{Traction bc collocation}
\end{equation}

\hspace{-0.23 in}The first Piola-Kirchhoff stress and undamaged free energy density fields are related to the displacement field through constitutive relations in equations \eqref{NeoHookean function}, \eqref{damage criterion}, \eqref{psio+ definition}, \eqref{damage irreversibility}, and \eqref{Cauchy stress neohookean}. Hence, the constitutive first Piola-Kirchhoff stress and undamaged free energy density fields are defined as 

\begin{equation}
\begin{split}
    &\tilde{\textbf{P}}^{\text{const}}(\textbf{X}^{i}_{d}, t^{j}, t^{j-1}) = \tilde{\textbf{P}}^{\text{const}}\Big(\hat{\textbf{u}}(\textbf{X}^{i}_{d}, t^{j}), \tilde{d}^{\text{const}}(\textbf{X}^{i}_{d}, t^{j-1})\Big), \\& \:\:\:\: \: \:\: \:\:\: \:\:\:\:\:\:\: \tilde{\psi}^{\text{const}}(\textbf{X}^{i}_{d}, t^{j}) = \tilde{\psi}^{\text{const}}\Big(\hat{\textbf{u}}(\textbf{X}^{i}_{d}, t^{j})\Big).
\label{P and psio constitutive}
\end{split}
\end{equation}

\hspace{-0.23 in}Note that $\tilde{d}^{\text{const}}$ is evaluated using $\tilde{\psi}^{\text{const}}$ and equations \eqref{damage criterion}, \eqref{damage irreversibility}. The constitutive relations can be applied through the following collocations

\begin{equation}
    \hat{\textbf{P}}(\textbf{X}^{i}_{d}, t^{j})  = \tilde{\textbf{P}}^{\text{const}}(\textbf{X}^{i}_{d}, t^{j}, t^{j-1}), \:\: \hat{\psi}(\textbf{X}^{i}_{d}, t^{j})  =  \tilde{\psi}^{\text{const}}(\textbf{X}^{i}_{d}, t^{j}).
\label{P and psio constitutive collocation}
\end{equation}

\hspace{-0.23 in}The total loss function $\mathcal{L}_{Total}(\boldsymbol{\theta}^{j})$ at an arbitrary time instant $t=t^{j}$ can now be constructed as the weighted mean squared error with contributions from the balances of linear and angular momentum, boundary conditions, and constitutive relations as

\begin{equation}
    \mathcal{L}_{Total} = \alpha_{BLM}\mathcal{L}_{BLM} + \alpha_{BAM}\mathcal{L}_{BAM} + \alpha_{BC}\mathcal{L}_{BC} + \alpha_{C}\mathcal{L}_{C}.
\label{loss function decomposition}
\end{equation}

\hspace{-0.23 in}$\mathcal{L}_{BLM}$, $\mathcal{L}_{BAM}$, $\mathcal{L}_{BC}$, $\mathcal{L}_{C}$ denote the individual loss functions for the balance of linear momentum, the balance of angular momentum, boundary conditions and constitutive relations, respectively. $\alpha$'s are the corresponding weights. During PINN training for a time instant $t^{j}$, the network parameters $\boldsymbol{\theta}^{j}$ are optimized to minimize the total loss function ($\mathcal{L}_{Total}$), 

\begin{equation}
    \breve{\boldsymbol{\theta}}^{j} = \underbrace{\text{argmin}}_{\boldsymbol{\theta}^{j}} \mathcal{L}_{Total}(\boldsymbol{\theta}^{j}).
\label{training network parameter optimization}
\end{equation}

\hspace{-0.23 in} The expanded forms of the individual loss functions that penalize errors in the collocation equations in the least-squares sense can be written as

\begin{equation}
    \mathcal{L}_{BLM} = \dfrac{\alpha_{BLM}}{N_{d}} \sum_{i=1}^{N_{d}} |\text{Div}\hat{\textbf{P}}(\textbf{X}^{i}_{d}, t^{j}, \boldsymbol{\theta}^{j}) - \rho_{R}\ddot{\hat{\textbf{u}}}(\textbf{X}^{i}_{d}, t^{j}, \boldsymbol{\theta}^{j})|^{2},
\label{least square BLM}
\end{equation}

\begin{equation}
\begin{split}
    \mathcal{L}_{BAM} &= \dfrac{\alpha_{BAM}}{N_{d}} \sum_{i=1}^{N_{d}} |\hat{\textbf{P}}(\textbf{X}^{i}_{d}, t^{j}, \boldsymbol{\theta}^{j}) \Big(\textbf{I}+\dfrac{\partial \hat{\textbf{u}}(\textbf{X}^{i}_{d}, t^{j}, \boldsymbol{\theta}^{j})}{\partial \textbf{X}}\Big)^{\text{T}} \\&- \Big(\textbf{I}+\dfrac{\partial \hat{\textbf{u}}(\textbf{X}^{i}_{d}, t^{j}, \boldsymbol{\theta}^{j})}{\partial \textbf{X}}\Big) \hat{\textbf{P}}^{\text{T}}(\textbf{X}^{i}_{d}, t^{j}, \boldsymbol{\theta}^{j})|^{2},
\label{least square BAM}
\end{split}
\end{equation}

\begin{equation}
\begin{split}
    \mathcal{L}_{BC} &= \alpha_{BCu} \mathcal{L}_{BCu} + \alpha_{BCt} \mathcal{L}_{BCt} \\&= \dfrac{\alpha_{u}}{N_{u}} \sum_{i=1}^{N_{u}} |\hat{\textbf{u}}(\textbf{X}^{i}_{u}, t^{j}, \boldsymbol{\theta}^{j}) - \overline{\textbf{u}}(\textbf{X}^{i}_{u}, t^{j})|^{2} \\&+ \dfrac{\alpha_{t}}{N_{t}} \sum_{i=1}^{N_{t}} |\hat{\textbf{P}}(\textbf{X}^{i}_{t}, t^{j}, \boldsymbol{\theta}^{j}) \textbf{n}_{R}(\textbf{X}^{i}_{t}) - \overline{\textbf{t}}_{R}(\textbf{X}^{i}_{t}, t^{j})|^{2},
\label{least square BC}
\end{split}
\end{equation}

\begin{equation}
\begin{split}
    \mathcal{L}_{C} &= \alpha_{S} \mathcal{L}_{S} + \alpha_{\text{UFE}} \mathcal{L}_{\text{UFE}} \\&= \dfrac{\alpha_{S}}{N_{d}} \sum_{i=1}^{N_{d}} |\hat{\textbf{P}}(\textbf{X}^{i}_{d}, t^{j}, \boldsymbol{\theta}^{j})  - \tilde{\textbf{P}}^{\text{const}}(\textbf{X}^{i}_{d}, t^{j}, t^{j-1}, \boldsymbol{\theta}^{j}, \breve{\boldsymbol{\theta}}^{j-1})|^{2} \\&+ \dfrac{\alpha_{\text{UFE}}}{N_{d}} \sum_{i=1}^{N_{d}} |\hat{\psi}(\textbf{X}^{i}_{d}, t^{j}, \boldsymbol{\theta}^{j})  - \tilde{\psi}^{\text{const}}(\textbf{X}^{i}_{d}, t^{j}, \boldsymbol{\theta}^{j})|^{2},
\label{least square C}
\end{split}
\end{equation}

\hspace{-0.23 in}with subscripts $S, \: \text{UFE}, \: BCu, \: BCt$ denoting terms associated with stress, undamaged free energy density, displacement boundary condition, and traction boundary conditions, respectively. $\tilde{d}$ from equation \eqref{PINN damage criterion} after training for every time step is stored in a global variable for use in the next time step's training calculations. The trained parameters $\breve{\boldsymbol{\theta}}$ are stored at desired intervals for visualizing the solution's temporal evolution. The approximate PINN solution for the damage problem at a referential point $\textbf{X}$ and time instant $t^{j}$ can then be evaluated for post-processing as

\begin{equation}
    (\hat{\textbf{u}}, \hat{\textbf{P}}, \hat{\psi}) = \mathcal{N}\mathcal{N}(\textbf{X}, t^{j}, \breve{\boldsymbol{\theta}}^{j}),
\label{Postprocessing at a point and time instant}
\end{equation}

\begin{equation*}
    \tilde{d}=\begin{cases}
\begin{aligned}
			&0 \hspace{0.2in} \hspace{0.1in} &&\text{if}  \hspace{0.1in} \hat{\psi}^{+} \leq \psi^{+}_{cr},\\
			&1  &&\text{if} \hspace{0.1in} \hat{\psi}^{+} > \psi^{+}_{cr},
\end{aligned}
	\end{cases}
\label{Postprocessing at a point and time instant 2} 
\end{equation*}

\begin{equation*}
    \tilde{\textbf{T}} = \tilde{J}^{-1} \hat{\textbf{P}} \tilde{\textbf{F}}^{\text{T}}, \: \tilde{\textbf{F}} =  \textbf{I}+\dfrac{\partial \hat{\textbf{u}}}{\partial \textbf{X}}, \: \tilde{J} = \text{det}\tilde{\textbf{F}}.
\label{Postprocessing at a point and time instant 3} 
\end{equation*}

\begin{figure*}[h!]
    \begin{center}
		\includegraphics[width=\textwidth]{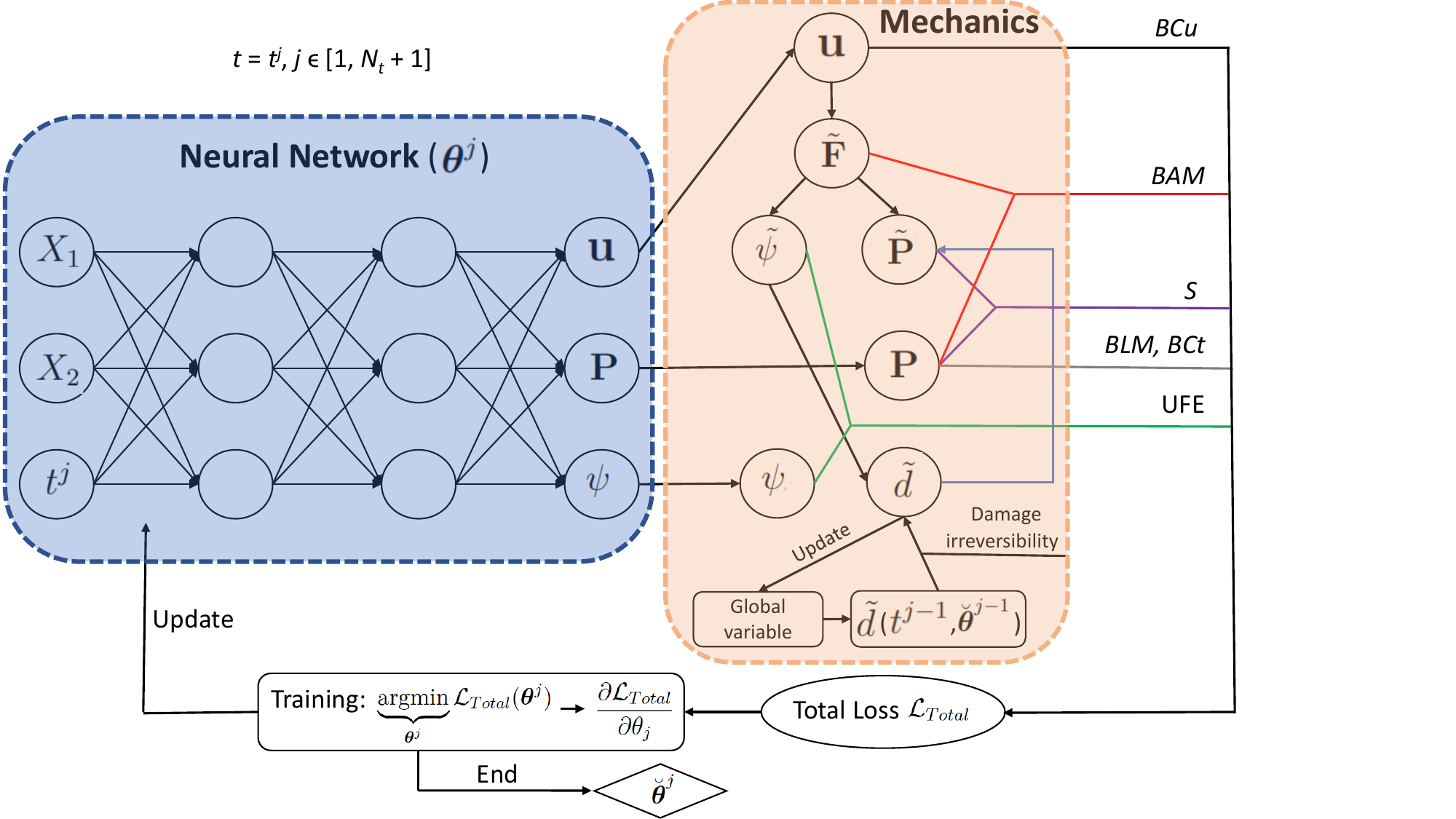}
	\end{center}
    \caption{{\textbf{The architecture of the PINN for large deformation fracture modeling.} The formulation is based on temporal discretization into $N_{t}$ time steps. The loss function formulation and training process for a time instant $t^{j}$, $j \in [1, N_{t}+1]$ is summarized. Mechanics-related information of boundary conditions ($BCu$, $BCt$), governing equations ($BLM$, $BAM$) and constitutive relations ($S$, UFE) is encoded in the total loss function $\mathcal{L}_{Total}$. $BLM$ and $BAM$ denote the individual loss functions corresponding to the balance of linear momentum and angular momentum, respectively. $BCu$ and $BCt$ are the individual loss functions corresponding to displacement and traction boundary conditions, respectively. $S$ and UFE are the individual loss functions for stress and undamaged free energy density constitutive relations, respectively.}} 
    \label{PDM-PINN architecture}
\end{figure*}

\hspace{-0.23 in}The architecture of the PINN for large deformation fracture modeling is summarized in Figure \ref{PDM-PINN architecture}.

For the fracture problems considered in this work, the inertial terms have a negligible effect, and they are neglected in equation \eqref{least square BLM} used for the PINN loss function formulation. In the baseline FEM solutions, we have incorporated the inertial terms and show that this simplification has negligible effects. The same neural network is trained for each time step without re-initialization, i.e., the trained network parameters for a time step are used as the initial values during training for the next time step. 

\section{Fracture predictions for various defect configurations}
\label{Numerical examples}

We apply the PINN to model fracture in three plane-strain geometries that capture a variety of defect (notches and holes) configurations. The chosen geometries probe different phenomena and are used in the literature to validate new numerical methods for modeling fracture. The units and material parameters are specified in a dimensionless manner, as is standard in PINN applications \citep{PINNdimensionless1, PINNdimensionless2}. The geometries are stretched uniaxially at a speed of 2 and have dimensions of $1 \times 1$. The values for $\mu$, $K$, and $\psi^{+}_{cr}$ are taken as 0.77, 1.66, and 3, respectively. Note that $\tilde{\textbf{P}}^{\text{const}}$ in equation \eqref{least square C} is obtained using $\tilde{\textbf{F}}$, equation \eqref{Cauchy stress neohookean} and the standard relation between $\textbf{P}$ and $\textbf{T}$ which encodes the balance of angular momentum as 

\begin{equation}
    \textbf{P} = J \textbf{T} \textbf{F}^{-\text{T}} \implies \tilde{\textbf{P}}^{\text{const}} = \tilde{J} \Big[\dfrac{\mu}{\tilde{J}}(\tilde{\textbf{B}}_{dis})_{0} + K(\tilde{J}-1)\textbf{I}\Big] \tilde{\textbf{F}}^{-\text{T}}, \: d=0.
\end{equation}

\hspace{-0.23 in}Hence, $\mathcal{L}_{BAM}$ in equation \eqref{loss function decomposition} is not considered in the total loss function for training. The default grids of collocation points for all geometries are obtained by creating a uniform $80 \times 80$ grid on the $1 \times 1$ referential domain and deleting the points lying inside the defects. In addition, the circular regions of the defect boundaries are seeded with 60 residual points. 

The PINN is implemented in the PyTorch framework and has three hidden layers, each with 30 neurons.  The loss weights $\alpha_{BLM}$, $\alpha_{C}$, $\alpha_{BC}$ are set equal to 1. A discussion on this choice can be found in the Appendix. The optimization algorithm used is `Adam', with an initial learning rate of 0.0001. Using an exponential decay scheduler, the learning rate is gradually decreased during training. The activation function is the ‘Tanh’ function, except for the last layer, where the ‘Softplus’ activation function is used. This configuration is shown to be optimal for capturing sharp interfaces in the computational domain \citep{SijunplasticityPINN}, here between the undamaged ($d=0$) and completely damaged ($d=1$) regions. Fourier random feature mapping is performed on the neural network input \textbf{X} = $\{X_{1}, X_{2}\}$ as

\begin{equation}
    \boldsymbol{\gamma}(\textbf{X}) = \Big(\text{cos}(\textbf{B}\textbf{X}), \text{sin}(\textbf{B}\textbf{X})\Big), 
\label{Fourier random feature mapping}
\end{equation}

\hspace{-0.23 in}with \textbf{B} being the Fourier random feature matrix with dimensions $m$ $\times$ 2. Entries in $\textbf{B}$ are sampled from a Gaussian distribution. $\boldsymbol{\gamma}(\textbf{X})$ is then used as an input to the neural network instead of \textbf{X}. This input mapping has been shown to help capture the high-frequency features in PINN solutions \citep{SijunplasticityPINN}, which are likely to exist here due to the sharp interfaces between undamaged and completely damaged regions. The training is performed with GPU acceleration on the computation cluster provided by the Center for Computation and Visualization (CCV). It is important to note that, due to hyperelasticity, training can be started at any time before damage initiation. This helps reduce the computation time. 

The finite element software ABAQUS/Explicit is used to implement FEM with gradient damage to obtain the baseline solutions for the PINN validation. ABAQUS has been used extensively to successfully model a broad variety of complex mechanics problems \citep{ABAQUSvaishakh, KonalePBS, Bai2021AEngineering, Zhong2021AGrowth, Srivastava2011STRESSPIPE}. The FEM with gradient damage implementation procedure using a combination of a user-defined material subroutine (VUMAT) for the deformation problem and a user-defined element (VUEL) for the damage problem described in \citep{KONALEMoM} is followed here. The gradient damage method's mathematical formulation is summarized as 

\begin{equation}
    \textbf{T} = g(d) \textbf{T}^{+}_{\text{o}}, \quad \textbf{T}_{\text{o}} = \dfrac{\mu}{J}(\textbf{B}_{dis})_{0} + K(J-1)\textbf{I},
\label{Cauchy stress GDM}
\end{equation}

\hspace{-0.23 in}with $g(d) = (1-d)^2$ being the degradation function and $\textbf{T}_{\text{o}}$, $\textbf{T}^{+}_{\text{o}}$ being the undamaged Cauchy stress and its distortional and tensile dilatational part, respectively. The evolution equation for $d$ is given as

\begin{equation}
    \zeta \dot{d} = 2(1-d)\mathcal{H} -2\psi_{*}(d- l^{2} \Delta d),
\label{Damage PDE GDM}
\end{equation}

\hspace{-0.23 in}where $\zeta>0$ is a constant kinetic modulus governing the timescale of damage growth. $\psi_{*}$ is a coefficient with units of energy per unit volume. It represents part of the energy dissipated during damage growth.  $\Delta$ denotes the Laplacian of a scalar field. $\mathcal{H}$ is a monotonically increasing history function expressed as
\begin{equation}
\begin{split}
    & \hspace{0 cm} \mathcal{H}(t) \overset{\text{def}}{=} \underset{s \in [0,t]}{\text{max}} \Big[\langle \psi^{+}(s) - \psi^{+}_{cr} \rangle\Big], \quad \psi = \dfrac{\mu}{2}(I_{1}-3) + \dfrac{K}{2}(J-1)^{2},
 \\& \langle \psi^{+}(s) - \psi^{+}_{cr} \rangle = \begin{cases}
\begin{aligned} 
			&0 \hspace{0.2in} \hspace{0.1in} &&\text{if}  \hspace{0.1in} \psi^{+}(s) - \psi^{+}_{cr} < 0,\\
			& \psi^{+}(s) - \psi^{+}_{cr}  && \text{if} \hspace{0.1in} \psi^{+}(s) - \psi^{+}_{cr} \geq 0. 
\end{aligned}
	\end{cases} 
\end{split}
\label{monotonically increasing history function}
\end{equation}

\hspace{-0.23 in}The number of elements used in the finite element meshes for each plate geometry is sufficiently large ($\sim$10,000) to ensure good accuracy of the baseline FEM with gradient damage solutions. Specifically, the finite element meshes used give results reasonably close to the converged solutions while not drastically increasing the computation time. The damage governing equations in the PINN [equations \eqref{damage criterion}, \eqref{damage irreversibility}], and the gradient damage method [equation \eqref{Damage PDE GDM}, \eqref{monotonically increasing history function}] are fundamentally different. Hence, the PINN fracture solutions are validated through comparison of the crack path with the baseline solutions. Point-wise and global $L_{2}$ errors are evaluated in the deformation stage before fracture. The normalized errors for displacement, Cauchy stress, and undamaged free energy density are defined as

\begin{equation}
    \text{Displacement}: L^{u}_{2} = \dfrac{||\hat{\textbf{u}}-\textbf{u}^{ref}||_{2}}{||\textbf{u}^{ref}||_{2}}, \: \textbf{u} = {u_{1}, u_{2}},
\label{Displacement L2 errors}
\end{equation}

\begin{equation}
    \text{Cauchy stress}: L^{S}_{2} = \dfrac{||\tilde{\textbf{T}}-\textbf{T}^{ref}||_{2}}{||\textbf{T}^{ref}||_{2}}, \: \tilde{\textbf{T}} = \{\tilde{{T}}_{11}, \tilde{{T}}_{22}, \tilde{{T}}_{12}\},
\label{Stress L2 errors}
\end{equation}

\begin{equation}
    \text{Undamaged free energy density}: L^{\text{UFE}}_{2} = \dfrac{||\hat{\psi}-\psi^{ref}||_{2}}{||\psi^{ref}||_{2}},
\label{Displacement L2 errors}
\end{equation}

\hspace{-0.23 in}where the superscript \emph{ref} denotes quantities obtained from the baseline FEM with gradient damage solutions. FEM without gradient damage solutions for different mesh structures were also obtained using ABAQUS/Explicit to reemphasize their mesh-dependency. For this, the deformation-only model implemented using a VUMAT was used along with the in-built element deletion in ABAQUS. The solutions of the PINN after training on the $80 \times 80$ grid were evaluated on a testing grid consisting of $1000 \times 1000$ points to test its extrapolation capabilities. For $L_{2}$ error evaluation and crack path comparison, nodal data exported from ABAQUS were interpolated on the testing grid. 

\subsection{Asymmetric double-edge notched plate}
\label{Asymmetric double-edge notched plate}

\begin{figure*}[h!]
    \begin{center}
		\includegraphics[width=0.9\textwidth]{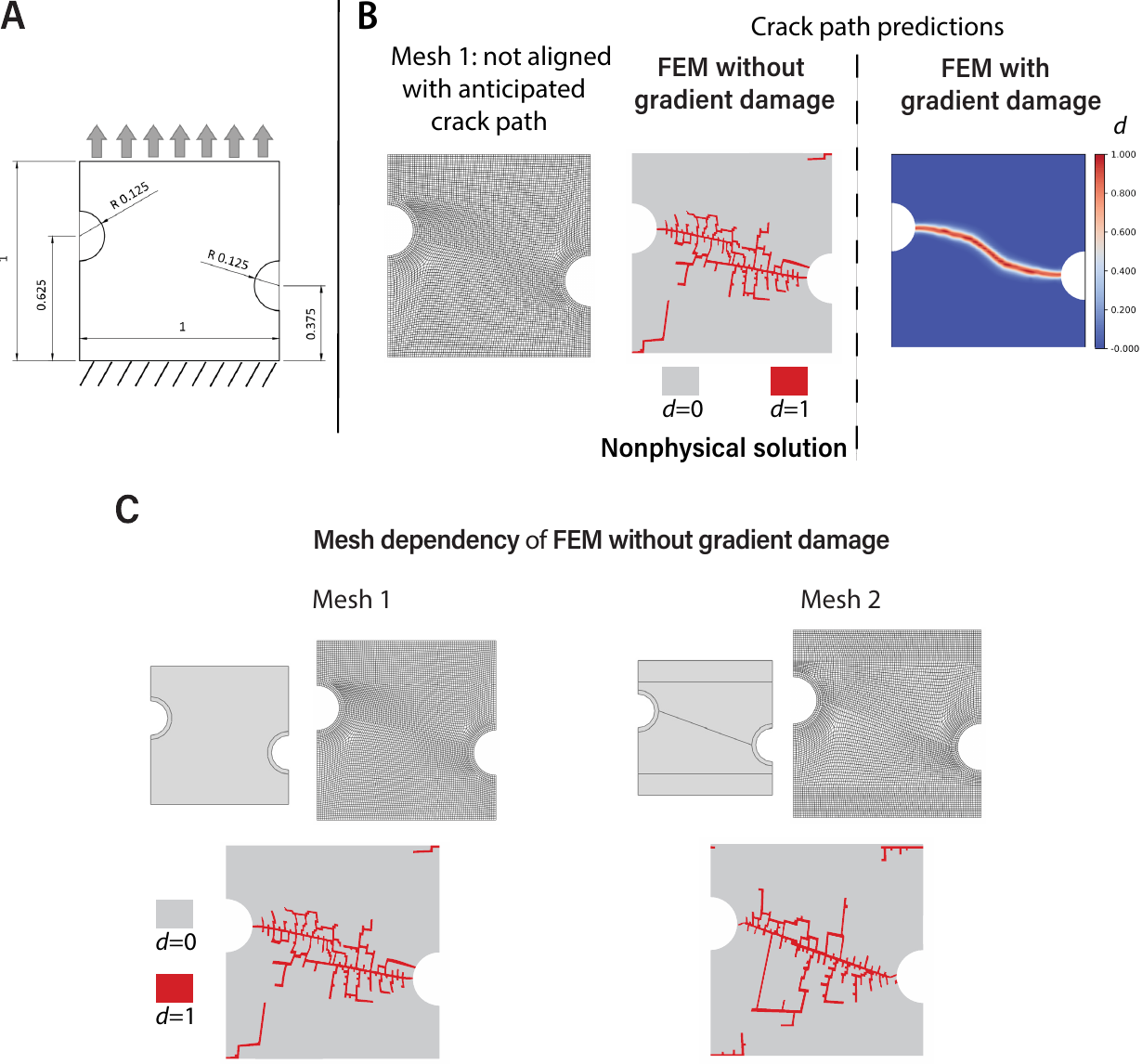}
	\end{center}
    \caption{{\textbf{Asymmetric double-edge notched plate: schematic, FEM with gradient damage and FEM without gradient damage crack path predictions.} \textbf{(A)} Schematic for an asymmetrically double-edge notched plate specimen subjected to tension under plane strain conditions. The geometry is stretched at a speed of 2. \textbf{(B)} Mesh not aligned with the anticipated crack path. The FEM without gradient damage solution, as expected, is nonphysical, while the FEM with gradient damage predicts the correct crack path. \textbf{(C)} The crack paths from FEM without gradient damage for two different meshes. The solutions differ significantly, highlighting FEM without gradient damage's mesh dependency.}} 
    \label{Two asymm holes FEM}
\end{figure*}

A specimen geometry often used to test the predictive capabilities of damage and fracture models in terms of crack path is a plate with asymmetric double-edge notches subjected to tension under plane-strain conditions \citep{Asymmdoublenotch, Asymmdoublenotch1, Asymmdoublenotch2}. The loading schematic and specimen geometry considered are shown in Figure \ref{Two asymm holes FEM}(A). The crack path predictions obtained using FEM with and without gradient damage for a mesh not aligned with the anticipated crack path are shown in Figure \ref{Two asymm holes FEM}(B). As expected, the results from FEM without a gradient damage solution are nonphysical. The significant mesh-dependency of FEM without gradient damage is shown through comparison of the predicted crack paths for two different meshes in Figure \ref{Two asymm holes FEM}(C). The nonphysical crack paths are dissimilar for the two meshes. In FEM without gradient damage, upon failure of an element, the subsequent damage growth depends on the local mesh structure around that element, resulting in the solution's strong mesh dependency.

\begin{figure*}[h!]
    \begin{center}
		\includegraphics[width=\textwidth]{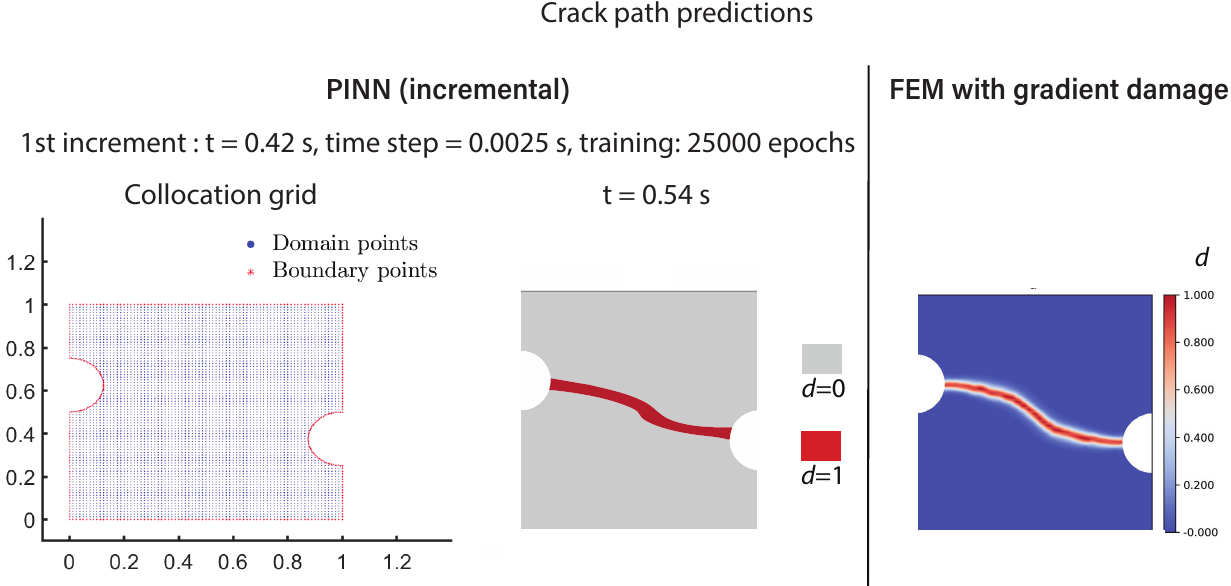}
	\end{center}
    \caption{{\textbf{Asymmetric double-edge notched plate: PINN (incremental) crack path prediction.} The collocation grid is shown. For the PINN (incremental) application, the time step size used is 0.0025 s, with the first increment being $t=0.42$ s. Training for each time increment is performed for 25,000 epochs. Damage solution from PINN (incremental) at $t=0.54$ s. The crack paths from PINN (incremental) and the baseline FEM with gradient damage show good agreement.}} 
    \label{Two asymm holes PDM-PINN (incremental)}
\end{figure*}

For the collocation grid in Figure \ref{Two asymm holes PDM-PINN (incremental)}, the PINN was applied incrementally [multiple time increments, referred to as PINN (incremental) hereon]. Specifically, the time step size was chosen to be 0.0025 s with the first increment at $t=0.42$ s. Training for each time increment was performed for 25,000 epochs. The time interval from the initiation of damage to the complete rupture ($t$ = 0.08 s) is small compared to the time to the initiation of damage (FEM with gradient damage: 0.477 s, PINN (incremental): 0.457 s), highlighting the rapid damage growth upon initiation built into the PINN formulation. The crack path predictions from the PINN (incremental) and the baseline FEM with gradient damage solution show good agreement, as shown in Figure \ref{Two asymm holes PDM-PINN (incremental)}.

\subsection{Single-edge notched plate with a hole}
\label{Single-edge notched plate with a hole}

\begin{figure*}[h!]
    \begin{center}
		\includegraphics[width=0.85\textwidth]{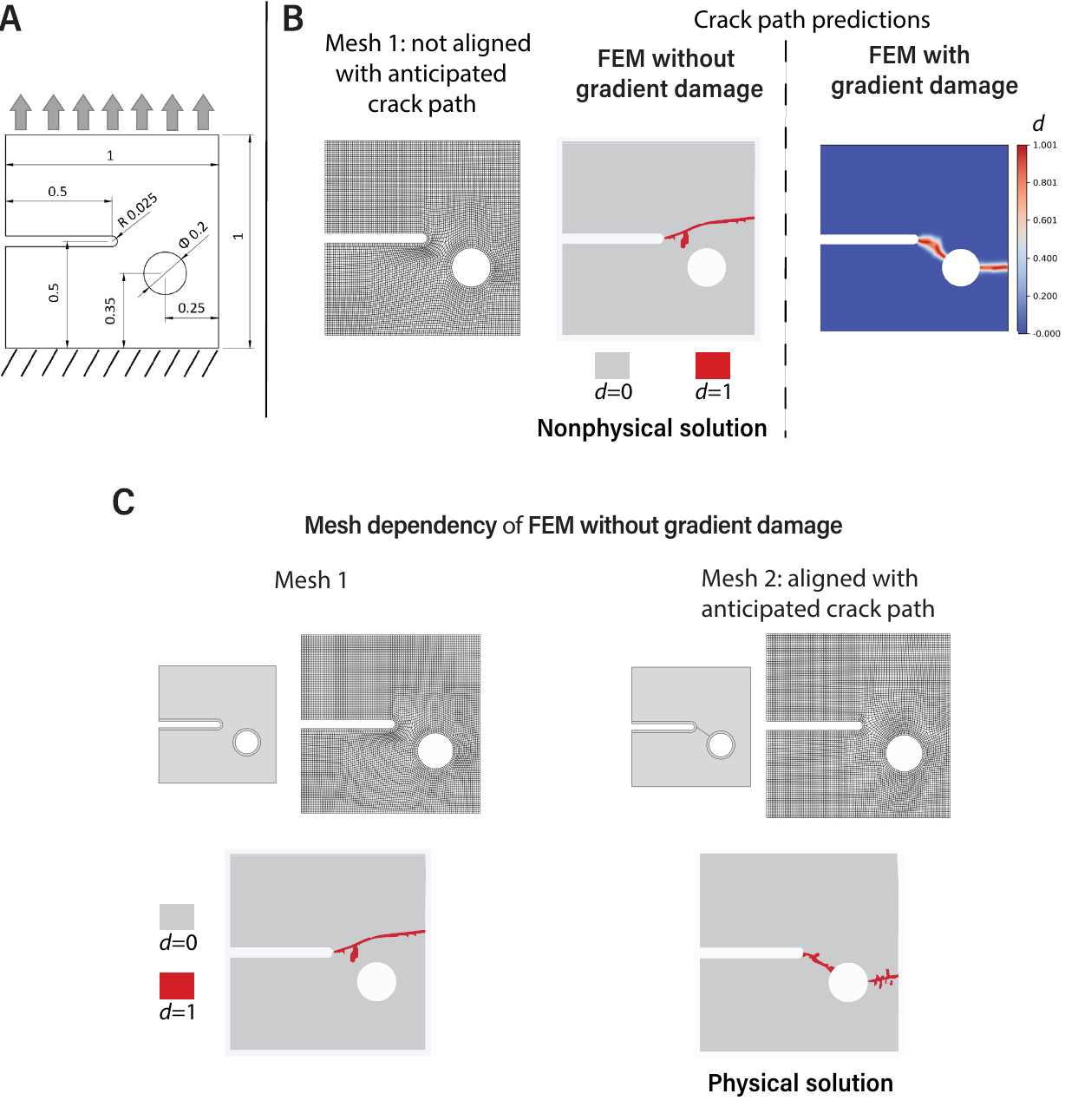}
	\end{center}
    \caption{{\textbf{Single-edge notched plate with a hole: schematic, FEM with and without gradient damage crack path predictions.} \textbf{(A)} Schematic for a single-edge notched plate with a hole subjected to tension under plane strain conditions. The geometry is stretched at a speed of 2. \textbf{(B)} The crack paths from FEM with and without gradient damage for a mesh not aligned with the anticipated crack path. The FEM without gradient damage solution is nonphysical. \textbf{(C)} Mesh dependency of FEM without gradient damage is highlighted, where a mesh aligned with the anticipated crack path shows the anticipated answer (shown on the right) while a general mesh without the knowledge of crack path makes a very incorrect prediction (shown on the left).}} 
    \label{notch and hole merging FEM}
\end{figure*}

A single-edge notched plate with a hole under plane-strain conditions is a specimen geometry utilized to validate damage and fracture models \citep{Notchandholemerging, Notchandholemerging1, Notchandholemerging2}. The loading schematic and specimen geometry considered are shown in Figure \ref{notch and hole merging FEM}(A). The crack path predictions obtained using FEM with and without gradient damage for a mesh not aligned with the anticipated crack path are shown in Figure \ref{notch and hole merging FEM}(B). The FEM without gradient damage solution is nonphysical. Figure \ref{notch and hole merging FEM}(C) shows the physical solution using FEM without gradient damage for a mesh aligned with the anticipated crack path, highlighting the method's mesh dependency.

\begin{figure*}[h!]
    \begin{center}
		\includegraphics[width=0.9\textwidth]{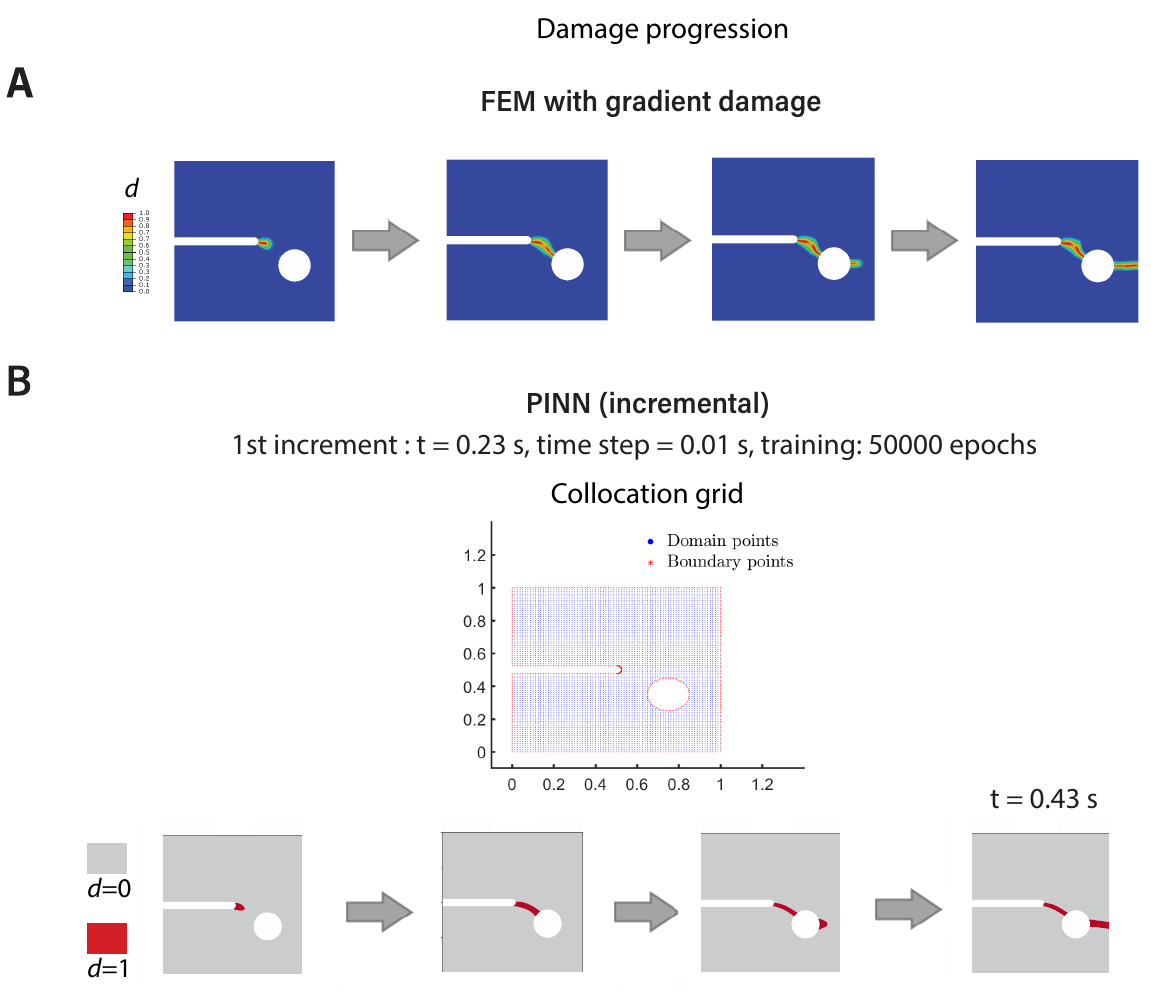}
	\end{center}
    \caption{{\textbf{Single-edge notched plate with a hole: PINN (incremental) damage progression and crack path predictions.} Damage progression at four different stages and crack paths from \textbf{(A)} FEM with gradient damage and \textbf{(B)} PINN (incremental) [along with collocation grid considered]. For the PINN (incremental) application, the time step size used is 0.01 s, with the first increment being $t=0.23$ s. Training for each time increment is performed for 50,000 epochs. Damage progression and crack paths from the PINN (incremental) and the FEM with gradient damage solutions show good agreement.}} 
    \label{notch and hole merging PDM-PINN (incremental)}
\end{figure*}

PINN (incremental) is applied to the collocation grid in Figure \ref{notch and hole merging PDM-PINN (incremental)}(B) with time step size = 0.01 s, the first time increment at $t=0.23$ s, and 50,000 epochs training for each time increment. The progression of damage at four different stages and the crack path in the PINN (incremental) solution agree with the FEM with gradient damage solution as seen in Figure \ref{notch and hole merging PDM-PINN (incremental)}.

\begin{figure*}[h!]
    \begin{center}
		\includegraphics[width=0.9\textwidth]{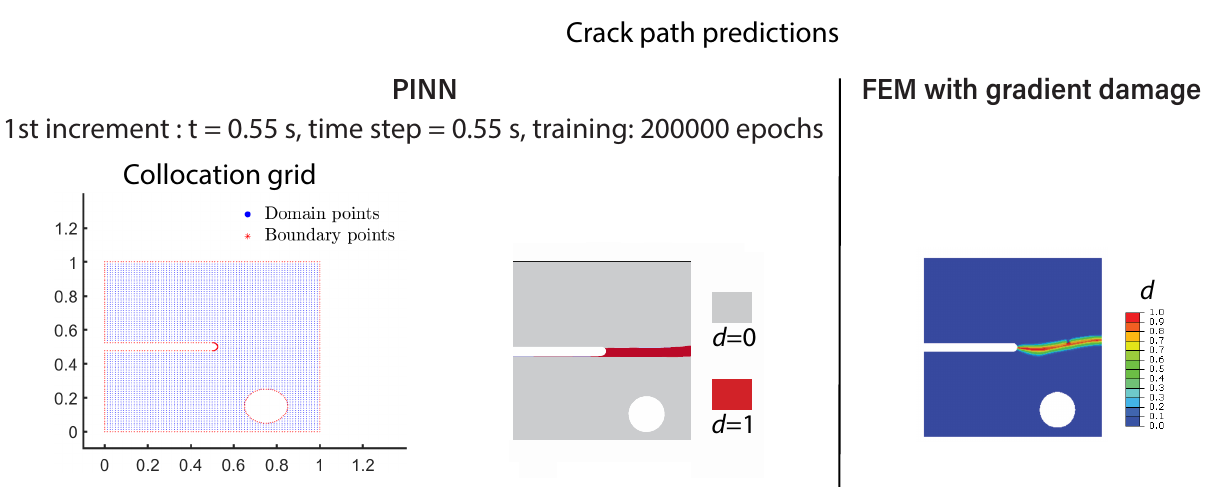}
	\end{center}
    \caption{{\textbf{Single-edge notched plate with a hole (increased notch and hole separation): PINN crack path prediction.} A variation of the single-edge notched plate with a hole geometry is considered. The separation of the hole center from the notch centerline in Figure \ref{notch and hole merging FEM}(A) is increased by 0.2. This results in a significant change in the crack path, as seen through the FEM with the gradient damage solution. The PINN solution was evaluated at $t=0.55$ s. The time step size used was 0.55 s, and training was performed for 200,000 epochs for the time increment. Good agreement of the crack path from the PINN with that from the FEM with baseline gradient damage solution. The PINN can accurately capture variations in the crack path with defect positioning.}} 
    \label{notch and hole merging far PDM-PINN}
\end{figure*}

We also consider a variation in this geometry where the hole is moved further away from the notch. Specifically, only the separation of the hole center from the notch centerline in Figure \ref{notch and hole merging FEM}(A) is increased by 0.2. The crack path in the FEM with gradient damage solution changes significantly as seen in Figure \ref{notch and hole merging far PDM-PINN}. The PINN [$t=0.55$ s, time step size = 0.55 s, 200,000 epochs training for the time increment] captures the crack path variation with the defect positioning accurately, as shown in Figure \ref{notch and hole merging far PDM-PINN}.

\subsection{Plate with four randomly distributed holes}
\label{Sheet with four randomly placed holes}

\begin{figure*}[h!]
    \begin{center}
		\includegraphics[width=0.9\textwidth]{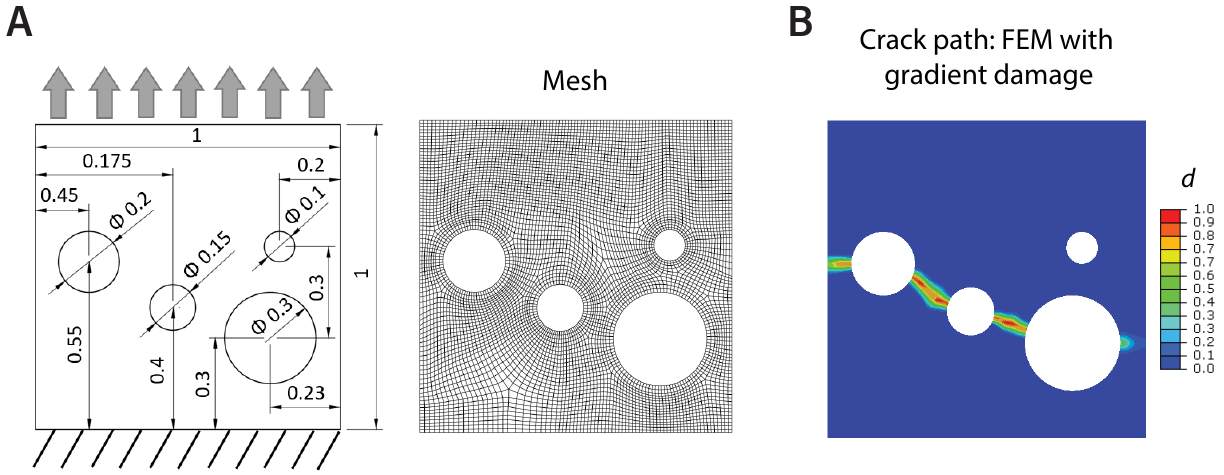}
	\end{center}
    \caption{{\textbf{Plate with four randomly distributed holes: schematic and FEM with gradient damage crack path prediction.} \textbf{(A)} Schematic for a plate with four randomly distributed holes subjected to tension under plane strain conditions. The geometry is stretched at a speed of 2. The mesh used for FEM with gradient damage application. \textbf{(B)} The crack path solution using FEM with gradient damage.}} 
    \label{four random asymm holes FEM}
\end{figure*}

We now consider a relatively complex geometry consisting of four holes with different diameters randomly distributed on the 1x1 plate domain. The loading schematic, specimen geometry, and finite element mesh considered are shown in Figure \ref{four random asymm holes FEM}(A). The crack path prediction obtained using FEM with gradient damage is shown in Figure \ref{four random asymm holes FEM}(B). 

\begin{figure*}[h!]
    \begin{center}
		\includegraphics[width=0.9\textwidth]{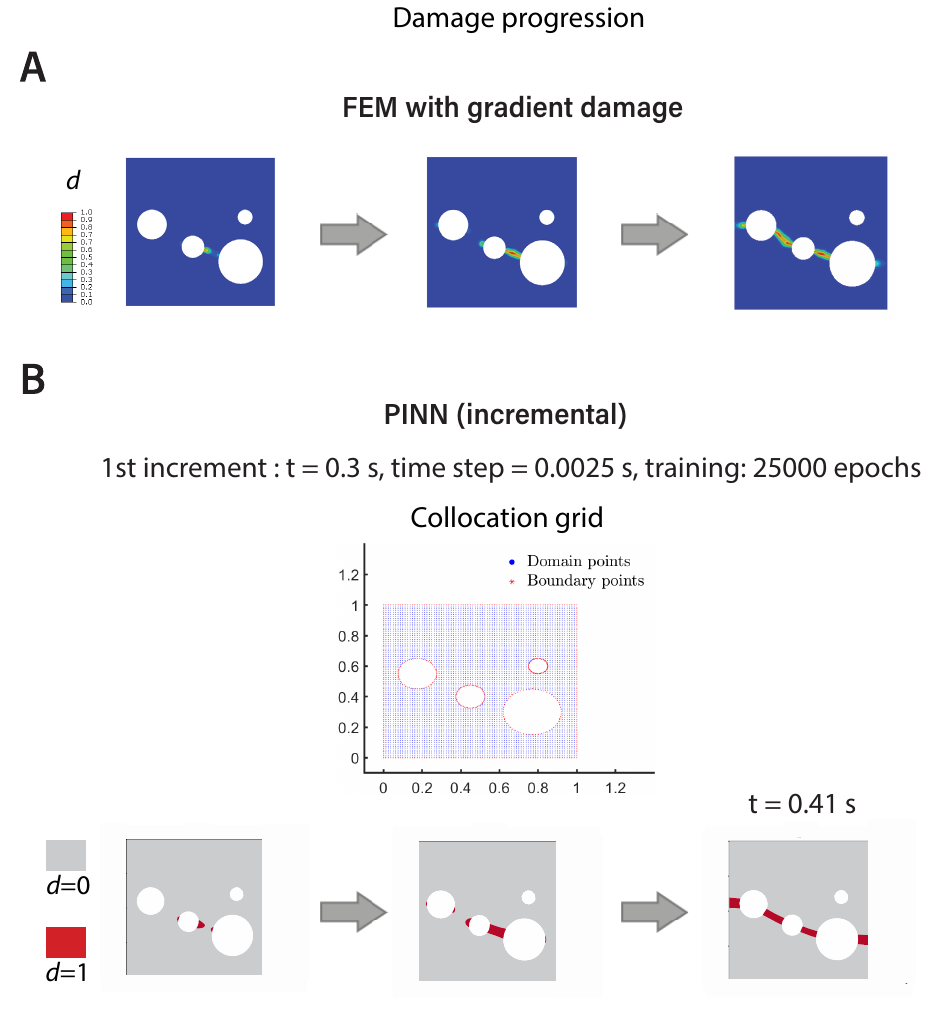}
	\end{center}
    \caption{{\textbf{Plate with four randomly distributed holes: PINN (incremental) damage progression and crack path predictions.} Damage progression at four different stages and crack paths from \textbf{(A)} FEM with gradient damage and \textbf{(B)} PINN (incremental) [along with collocation grid considered]. For the PINN (incremental) application, the time step size used is 0.0025 s, with the first increment being $t=0.3$ s. Training for each time increment is performed for 25,000 epochs. Good agreement of the damage progression and crack paths from the PINN (incremental) and the FEM with gradient damage solutions.}} 
    \label{four random asymm holes PDM-PINN (incremental)}
\end{figure*}

The PINN (incremental) is applied to the collocation grid in Figure \ref{four random asymm holes PDM-PINN (incremental)}(B) with time step size = 0.0025 s, the first time increment at $t=0.3$ s, and 25,000 epochs training for each time increment. Good agreement of the damage progression at four different stages and the crack path from the PINN (incremental) solution with the baseline FEM with gradient damage solution can be seen in Figure \ref{four random asymm holes PDM-PINN (incremental)}.

\section{Assessment of the PINN performance for fracture}
Previously, we demonstrated the success of PINN without gradient damage formulation in predicting the fracture path for a variety of problems. The performance of any new and emerging computational method has to be evaluated to guide the choice of underlying parameters for future applications. Towards this, we investigate the effect of a few key neural network parameters on the PINN's prediction performance. The PINN solutions in this section for the asymmetric double-edge notched plate geometry are obtained using a relatively computationally inexpensive single time increment approach [referred to as PINN (one-time step) hereon]. In this work, we have considered (i) hyperelastic (rate-independent) material response, (ii) instantaneous damage growth, which is approximately observed in experiments for elastomers and several soft polymers, where damage, when it initiates, grows very rapidly, and (iii) monotonic loading. Under these conditions, the PINN (one-time step) solutions provide crack paths with \emph{reasonable} accuracy, as shown in Appendix Figure A\ref{two asymm holes incremental one step comp}. For the general class of fracture problems, PINN (incremental) [multiple time increments] solutions will have better accuracy. However, PINN (one-time step) can enable computationally cheaper neural network parameter and architecture optimization before PINN (incremental) is applied for accurate final results.  

The crack paths are obtained through PINN (one-time step) solutions at $t=0.65$ s with 100,000 epochs of training. $L_{2}$ relative errors and losses are evaluated through PINN (one-time step) application at $t=0.45$ s (before damage initiation) with 100,000 epochs training. First, variations in the collocation point distribution (approximately constant grid density) are considered. Further, the number of layers, the number of neurons in each layer, and the grid density (in a fixed grid structure) are systematically varied. The base number of layers, number of neurons in each layer, and grid density are the values used in Section \ref{Numerical examples}, i.e., $5$, $30$, and $80 \times 80$, respectively. We also compare the computation time for the PINN using the base network and base grid density with those for FEM without and with gradient damage. The effects of simple pre-training and output normalization strategies on the solutions are also explored. 

\subsection{$L_{2}$ relative error and loss evolution during training}

\begin{figure*}[h!]
    \begin{center}
		\includegraphics[width=0.9\textwidth]{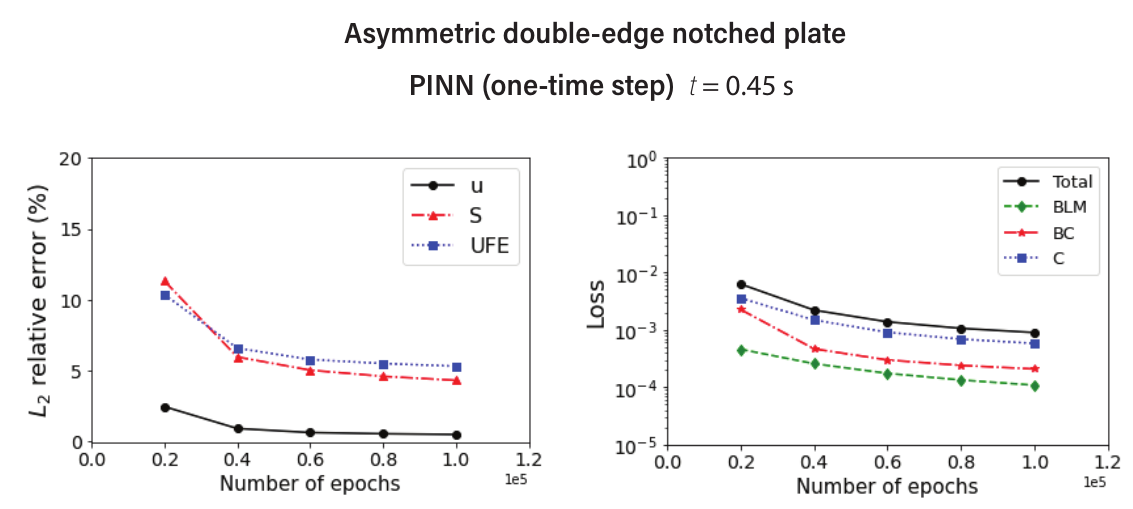}
	\end{center}
    \caption{{\textbf{Asymmetric double-edge notched plate: $L_{2}$ relative error and loss evolution during training.} PINN (one-time step) is applied at $t=0.45$ s (before damage initiation) with 100,000 epochs training. The following notations are used: $u$ - displacement, $S$ - stress, UFE - undamaged fee energy, $BLM$ - balance of linear momentum, $BC$ - boundary conditions, $C$ - constitutive equations. A monotonic decrease in errors and losses is observed during training.}} 
    \label{two asymm holes PDM-PINN (one time step) error and loss}
\end{figure*}

$L_{2}$ relative error ($u$ - displacement, $S$ - stress, UFE - undamaged free energy density) and loss (total, $BLM$ - balance of linear momentum, $BC$ - boundary conditions, $C$ - constitutive equations) evolution with number of epochs during training [PINN (one-time step), $t$=0.45 s, 100,000 epochs, grid from Figure \ref{Two asymm holes PDM-PINN (incremental)}] for the asymmetric double-edge notched plate geometry is shown in Figure \ref{two asymm holes PDM-PINN (one time step) error and loss}. The errors and losses decrease monotonically with the number of epochs in this range. Corresponding results for the single-edge notched plate with a hole and the plate with four randomly distributed holes geometries are presented in the Appendix. The low errors with respect to the baseline FEM solutions highlight the validity of the simplification discussed in Section \ref{lossfunctionconstruction}. 

\subsection{Effect of collocation point distribution}

\begin{figure*}[h!]
    \begin{center}
		\includegraphics[width=0.75\textwidth]{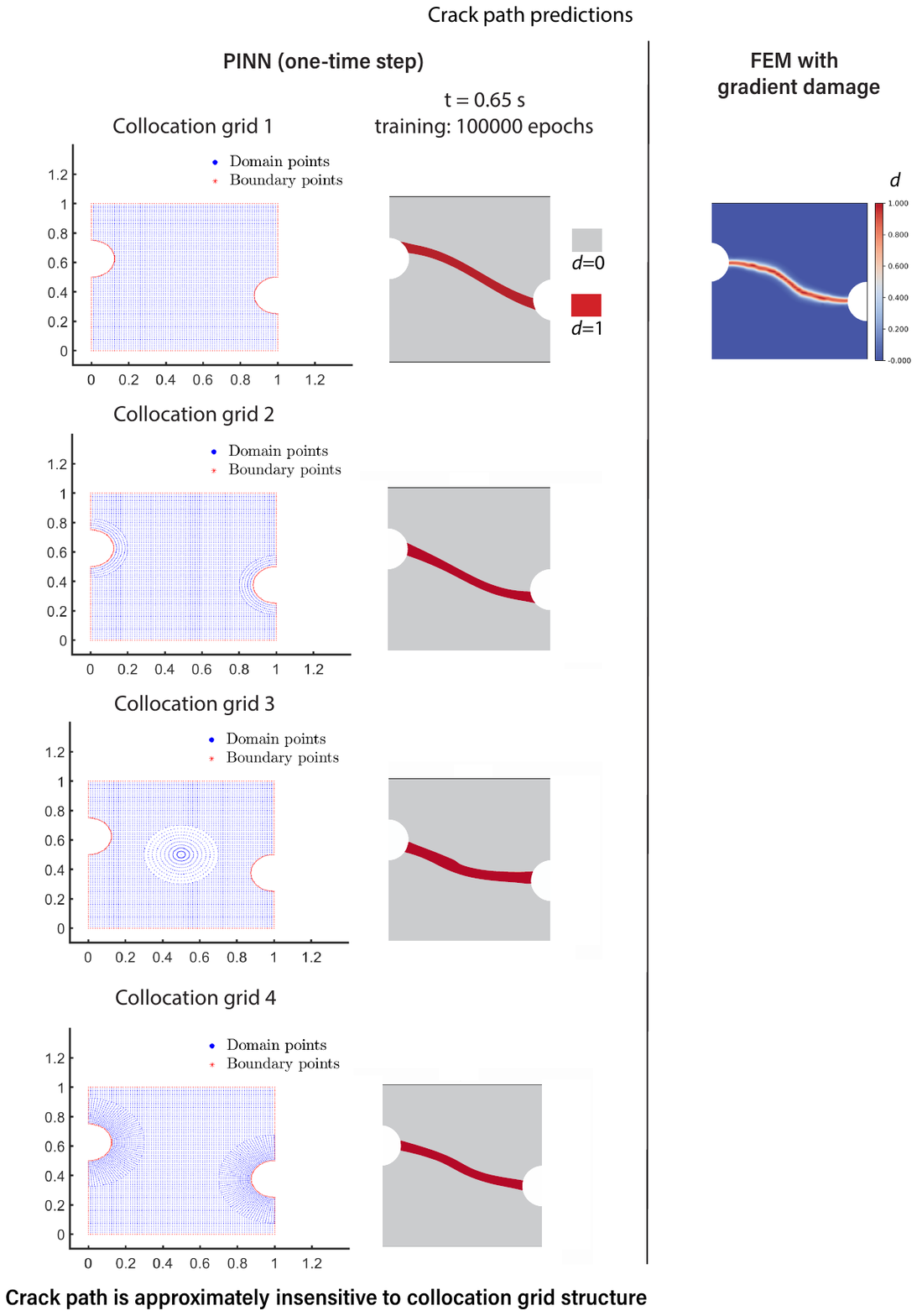}
	\end{center}
    \caption{{\textbf{Asymmetric double-edge notched plate: effect of collocation point distribution on crack path.} The four different collocation point distributions considered are shown. PINN (one-time step) is applied at $t=0.65$ s with training for 100,000 epochs. The crack paths for the four distributions using PINN (one-time step). The crack paths are approximately insensitive to the collocation point distribution, contrary to the significant mesh-dependency of FEM without gradient damage solutions (see Figure \ref{Two asymm holes FEM}). Good agreement of the PINN (one-time step) crack path predictions for the four distributions with the baseline FEM with gradient damage solution.}} 
    \label{Two asymm holes PDM-PINN (one time step)}
\end{figure*}

The four different collocation point distributions considered and the corresponding PINN (one-time step) solutions for the asymmetric double-edge notched plate geometry are shown in Figure \ref{Two asymm holes PDM-PINN (one time step)}. The PINN crack path solutions are approximately insensitive to the distribution and show good agreement with the baseline FEM with gradient damage solution. Corresponding results for the single-edge notched plate with a hole and the plate with four randomly distributed holes geometries are presented in the Appendix. 

The length scale parameter in FEM with gradient damage creates a diffuse damage zone around a completely damaged element. The restriction of $h_{e} \lesssim 0.2 l$ (typical element size) prescribed for FEM with gradient damage in the literature ensures that the diffuse zone is sufficiently large for the local mesh structure not to affect the following damage growth and the crack path. The meshless form of the PINN, i.e., physics-informed loss function minimization at collocation points not connected by a mesh, results in the crack path solutions being approximately insensitive to the collocation point distribution.  

\subsection{Effect of the number of layers, number of neurons in each layer, and grid density}

\label{factorial experiment}

The three different numbers of layers, number of neurons in each layer, and grid density considered were 3, 5, 8; 10, 30, 50; and $20 \times 20$, $40 \times 40$, $80 \times 80$, respectively. A numerical $3 \times 3 \times 3$ Factorial Design of Experiments (DOE) was performed to understand the main, simple, and interacting effects of the three hyperparameters. The complete results for the $L_{2}$ relative errors and the crack paths for the DOE study are presented in the Appendix. Figures \ref{two asymm holes PDM-PINN (one time step) factorial bf main text}(A), (B), (C) and (D) show the $L_{2}^{\text{UFE}}$ \% errors for a few cases, highlighting the effect of (i) number of layers ($80 \times 80$ grid density), (ii) number of neurons in each layer ($80 \times 80$ grid density),(iii) grid density (5 layers), and (iv) grid density (30 neurons in each layer), respectively. All three factors have main effects. Significant interaction effects are observed between the number of layers and the number of neurons, as well as between the grid density and the number of layers. Underfitting or overfitting occurs when the neural network parameters are not optimal (see Figures A\ref{two asymm holes PDM-PINN (one time step) factorial bf} and A\ref{two asymm holes PDM-PINN (one time step) factorial af}).

\begin{figure*}[h!]
    \begin{center}
		\includegraphics[width=\textwidth]{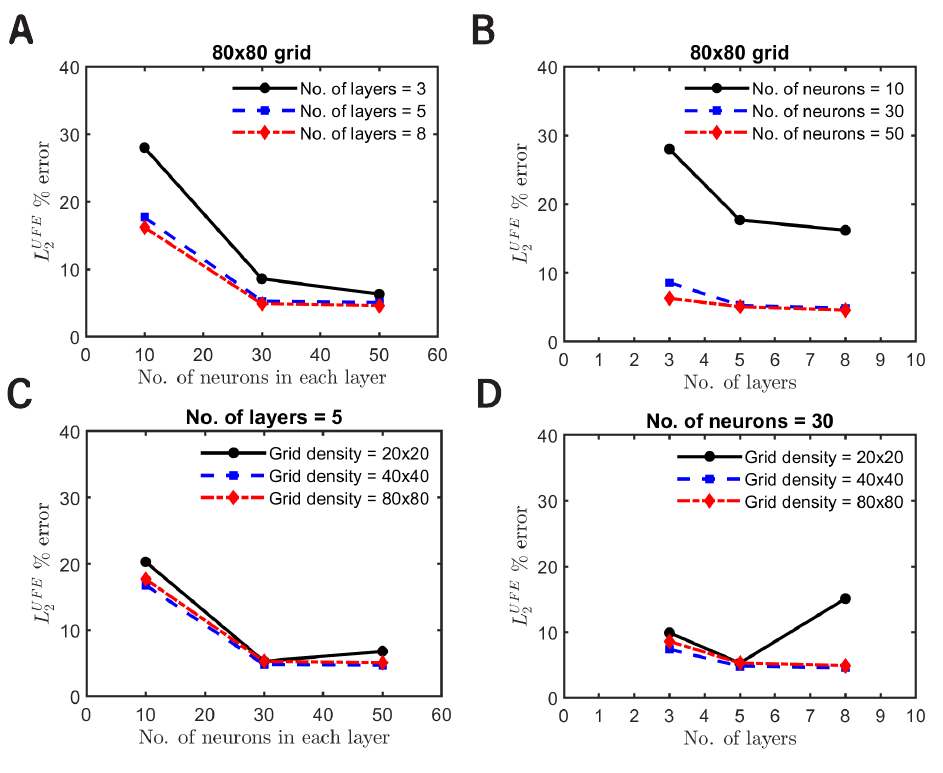}
	\end{center}
    \caption{{\textbf{Asymmetric double-edge notched plate: numerical factorial experiment ($L_{2}^{\text{UFE}}$ \% errors, $t$ = 0.45 s).} The complete results are presented in the Appendix. For $L_{2}$ error comparison, PINN (one-time step) is applied at $t=0.45$ s with 100,000 epochs training. The effects of (\textbf{A}) the number of layers ($80 \times 80$ grid density), (\textbf{B}) the number of neurons in each layer ($80 \times 80$ grid density), (\textbf{C}) grid density (5 layers) and \textbf{(D)} grid density (30 neurons). The three factors each exhibit a main effect. Significant interaction effects can be seen between the number of layers and the number of neurons, and between grid density and the number of layers. Non-optimal hyperparameter values lead to underfitting or overfitting.}} 
    \label{two asymm holes PDM-PINN (one time step) factorial bf main text}
\end{figure*}

\subsection{Comparison of computation time: PINN, FEM without and with gradient damage}

The computation time using the base network (5 layers each with 30 neurons) and the base grid density ($80 \times 80$) for 1 epoch of PINN training remained approximately constant for all numerical problems considered and was equal to 0.06 s (GPU accelerated, QuadroRTX, CCV@Brown), 0.18 s (GPU accelerated, NVIDIA T600 \& Intel Core i7-10700 Processor, Dell Precision 3650 Tower), and 0.24 s (not GPU accelerated, CCV@Brown). FEM without and with gradient damage (both implemented in ABAQUS/Explicit (2018)) were parallelized using 32 processors on a system with an Intel Xeon Gold 6326 CPU. For the finite element meshes considered in this work ($\sim$10,000 elements), the approximate computation time for an increment with a time step size of approximately $3 \:\: \text{x} \:\: 10^{-7}$ s was 0.003 s for both FEM approaches.  

PINN (incremental) in general requires more computation time than FEM. The complexity of computationally implementing gradient damage models in FEM and the associated implementation time costs are important considerations. Some reduction in computation time for PINN (incremental) can be achieved when the material response is rate-independent. As done in this work for an elastomer with hyperelastic response, the solution evaluation in PINN (incremental) for a rate-independent material can be started at any time instant before damage initiation. The computation time for PINN (one-time step), however, can be comparable to FEM.

\subsection{Effect of pre-training}

\begin{figure*}[h!]
    \begin{center}
		\includegraphics[width=0.72\textwidth]{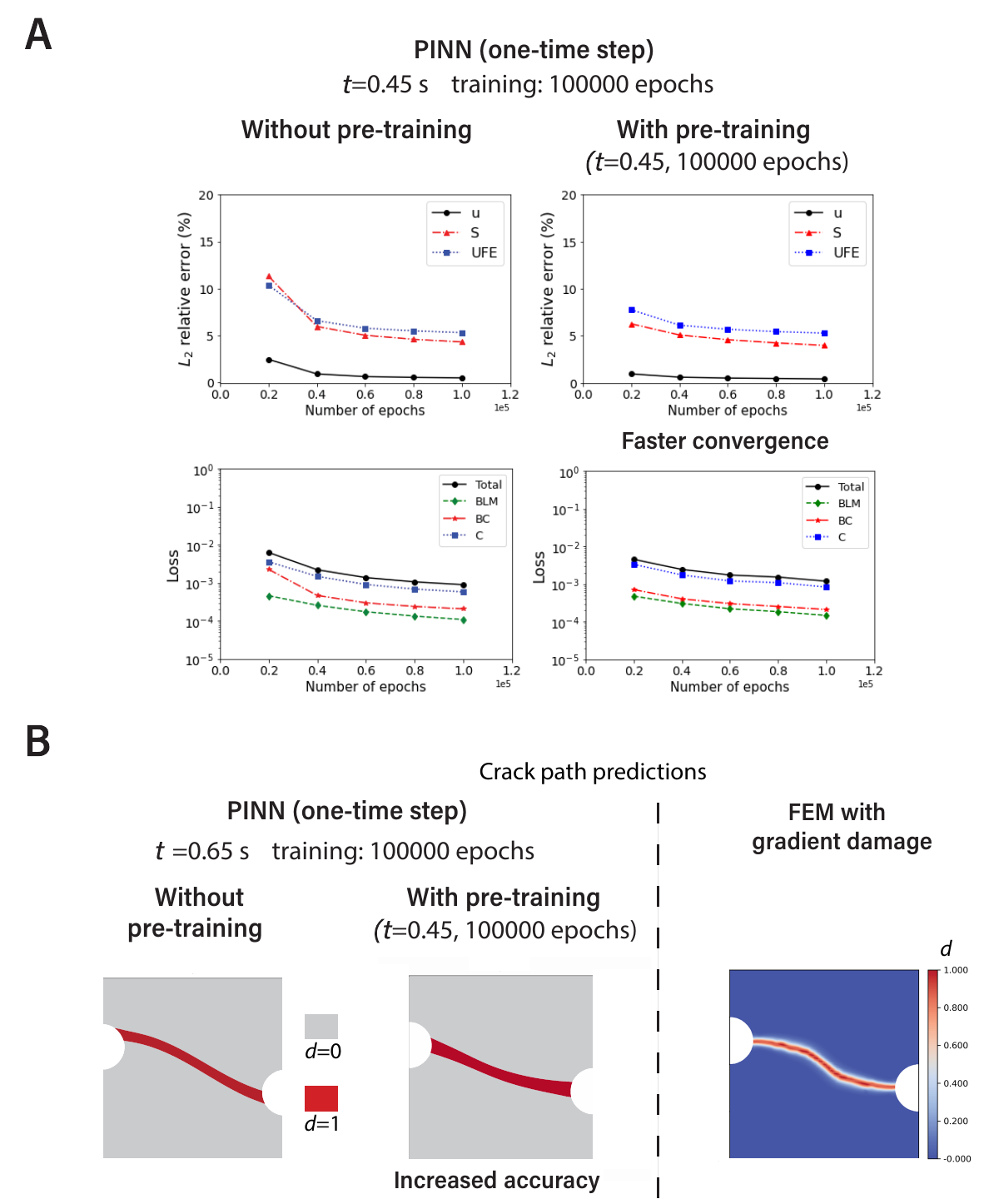}
	\end{center}
    \caption{{\textbf{Asymmetric double-edge notched plate: effect of pre-training.} The same material properties and boundary conditions discussed in Section \ref{Numerical examples} were used for pre-training on the 1x1 domain with a uniform 80 x 80 collocation point grid. \textbf{(A)} Comparison of $L_{2}$ errors and losses from PINN (one-time step) application at $t=0.45$ s, 100,000 training epochs with and without pre-training ($t=0.45$ s, 100,000 epochs). Faster convergence is achieved with pre-training. \textbf{(B)} Comparison of crack paths from PINN (one-time step) application at $t=0.65$ s, 100,000 epochs with and without pre-training ($t=0.45$ s, 100,000 epochs) and from the baseline FEM with gradient damage solution. The simple pre-training increases the accuracy of the solution.}}
    \label{pretraining figure}
\end{figure*}

Pre-training has been shown in the literature to improve the accuracy and convergence of PINNs \citep{Pretrainingone, Pretrainingtwo}. We investigate the effect of one pre-training strategy. Specifically, the $1 \times 1$ domain seeded with a uniform 80 x 80 collocation point grid is assigned the same material properties and boundary conditions discussed in Section \ref{Numerical examples}. The trained network parameters from stretching this unit square ($t$ = 0.45 s in one-time step, 100,000 epochs) are used to initialize the PINN for application to the asymmetric double-edge notch plate geometry. Comparison between the PINN (one-time step) results (at $t$ = 0.45 s [before damage initiation] and 0.65 s [after damage initiation], 100,000 epochs) with and without pre-training in Figure \ref{pretraining figure} shows the simple pre-training associated convergence and accuracy enhancement.

\subsection{Effect of PINN output normalization}

\begin{figure*}[h!]
    \begin{center}
		\includegraphics[width=0.72\textwidth]{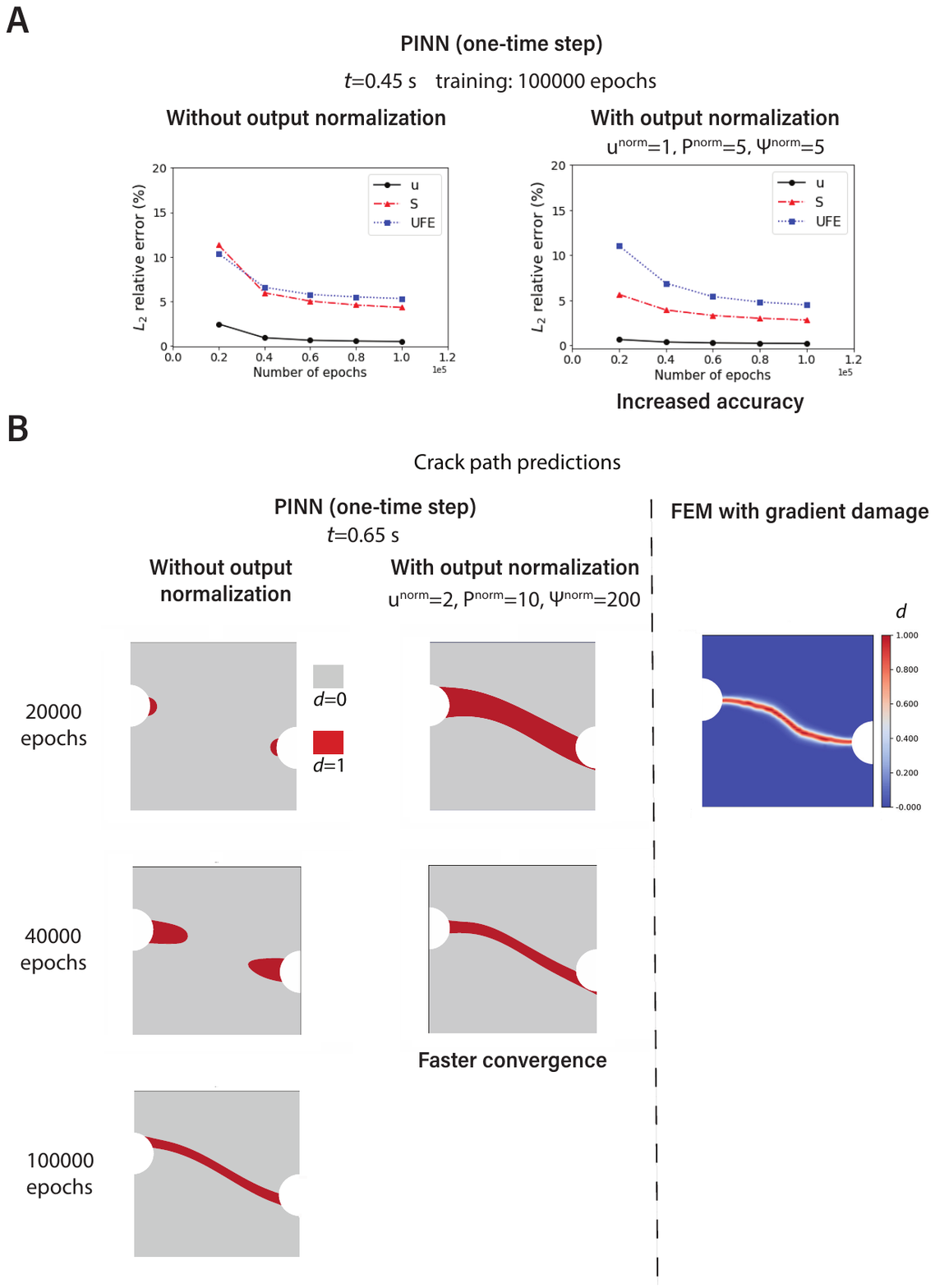}
	\end{center}
    \caption{{\textbf{Asymmetric double-edge notched plate: effect of output normalization.} \textbf{(A)} Comparison of $L_{2}$ errors from PINN (one-time step) application at $t=0.45$ s, 100,000 training epochs with and without output normalization ($u^{norm}=1$, $P^{norm}=5$, $\psi^{norm}=5$). Faster convergence and increased accuracy are achieved with output normalization. \textbf{(B)} Comparison of crack paths from PINN (one-time step) application at $t=0.65$ s, 100,000 epochs with and without output normalization ($u^{norm}=2$, $P^{norm}=10$, $\psi^{norm} = 200$) and from the baseline FEM with gradient damage solution. The simple output normalization led to increased accuracy of the solution and faster convergence.}}
    \label{output normalization figure}
\end{figure*}

Normalization of PINN outputs is another technique that has been shown in the literature to enhance accuracy and convergence \citep{Outputnormalization1, Outputnormalization2}. We consider a simple normalization scheme in the form of fixed normalization numbers that ensures physical consistency. Specifically, $\{\hat{\textbf{u}}, \overline{\textbf{u}}\}, \{\hat{\textbf{P}}, \tilde{\textbf{P}}^{const}, \overline{\textbf{t}}_{R}\}, \{\hat{{\psi}}, \tilde{{\psi}}^{const} \}$ are normalized by $u^{norm}, P^{norm}, \psi^{norm}$, respectively before evaluating the loss functions in equations \eqref{least square BLM}, \eqref{least square BC}, \eqref{least square C}. Figure \ref{output normalization figure} shows the comparison between the PINN (one-time step) results (at $t$ = 0.45 s [before damage initiation] - $u^{norm}=1$, $P^{norm}=5$, $\psi^{norm}=5$, 100,000 epochs and 0.65 s [after damage initiation] - $u^{norm}=2$, $P^{norm}=10$, $\psi^{norm} = 200$, 100,000 epochs) for the asymmetric double-edge notched plate geometry with and without output normalization. Faster convergence and increased accuracy are achieved with this simple output normalization scheme. 

\section{Conclusions}

The important and evolving field of fracture modeling commonly uses the finite element method (FEM). FEM without gradient damage is relatively easy to implement. However, the corresponding results are highly mesh-dependent and often inaccurate. FEM with gradient damage enables mesh-independent predictions at the cost of increased mathematical and numerical implementation complexities. Specifically, additional spatial gradient calculations and complex user-defined element and user subroutine development are necessary. Physics-informed neural networks (PINNs) have recently emerged as a computational method that can be applied to solve solid mechanics problems. The limited studies on PINNs for damage and fracture modeling in the literature are confined to the small-strain regime and use the gradient damage formulation by default. Due to the meshless nature of PINNs, we posed the question of whether we can correctly solve fracture problems without needing the gradient damage formulation. We propose a PINN without gradient damage to model large deformation fracture in elastomers. The PINN does not require training data and utilizes the collocation method. The capability of the PINN to accurately predict the crack path has been validated by studying a variety of benchmark defect configuration problems, where we used accurate solutions from FEM with gradient damage and a highly refined mesh. We also conducted an assessment of the PINN as a fracture modeling method by systematically varying a few key parameters to guide future design of PINNs for fracture problems. The PINN crack path predictions are largely insensitive to the collocation point distribution. The proposed modeling framework and neural network architecture could motivate the application of PINNs without gradient damage for modeling fracture in a broad class of materials. 

The relatively new nature of PINNs as a numerical method implies that the solution quality depends on a range of factors, and problem-dependent insights are necessary for the successful application of PINNs. However, the relative ease of mathematical formulation, numerical implementation, and mesh-free nature of PINNs are significant advantages. Future work can utilize adaptive collocation point distribution and explore new pre-training and output normalization strategies. The PINN architecture and implementation can be made more computationally efficient to allow for very small time step sizes and training for a higher number of epochs.

\section*{Author Contributions}
\textbf{Aditya Konale}: Investigation, Methodology, Formal analysis, Data curation, Software, Visualization, Writing – original draft preparation, review \& editing.  \textbf{Vikas Srivastava}: Conceptualization, Investigation, Methodology, Supervision, Funding acquisition, Writing – original draft preparation, review \& editing.

\section*{Acknowledgments}
The authors thank Dr. Khemraj Shukla at Brown University for discussions on the topic. The authors would like to acknowledge and thank the support from the Office of Naval Research (ONR), USA, to V. Srivastava under grants N00014-21-1-2815 and N00014-23-1-2688.

\section*{Declaration of Interest}
The authors declare no competing interests.

\begin{appendices}

\clearpage

\section*{Appendix}

\hspace{-0.55 cm}\underline{Asymmetric double-edge notched plate}

\subsection*{Effect of loss weights}

The individual loss terms (using all weights = 1) for the asymmetric double-edge notched plate ($t=0.45$ s, 100,000 training epochs) are not in the same order of magnitude as seen in Figure \ref{two asymm holes PDM-PINN (one time step) error and loss}. Particularly, the constitutive loss $\mathcal{L}_{C}$ is significantly higher than the balance of linear momentum (BLM) and the boundary condition (BC) losses. The loss weights $\alpha$ can be adjusted to further minimize the total loss. First, only $\alpha_{C}$ was increased (= 10, 100) to reduce $\mathcal{L}_{C}$. Figure A\ref{lossweighteffect}(A) shows the corresponding loss and $L_{2}$ error evolution with training epochs. $\mathcal{L}_{C}$ decreased with increasing $\alpha_{C}$ as anticipated. However, both $\mathcal{L}_{BC}$ and $\mathcal{L}_{BLM}$ increased, with a significant increase in $\mathcal{L}_{BC}$. The total loss also increased along with the $L_{2}$ errors. Since $\mathcal{L}_{BC}$ was the most sensitive to an increase in $\alpha_{C}$, we also simultaneously increased $\alpha_{BC}$ ($\alpha_{C}$ = $\alpha_{BC}$ = 10 and 100). A similar observation was made with $\mathcal{L}_{BLM}$ increasing with the weights, and hence the total loss increasing as seen in Figure A\ref{lossweighteffect}(B). The $L_{2}$ errors changed insignificantly compared to the baseline (all weights = 1) in this case. Finally, the three weights were simultaneously increased ($\alpha_{C}$ = $\alpha_{BC}$ = $\alpha_{BLM}$ = 10 and 100). The corresponding losses and $L_{2}$ error evolution were similar to the baseline, as shown in A\ref{lossweighteffect}(C). Hence, it was concluded that the baseline weight values provide a good balance between loss and $L_{2}$ error minimization. It can be noted that the optimal loss weight values can vary for different sets of material parameters (e.g., elastic modulus orders of magnitude different from that used in this study). Suitable weights need to be obtained for specific boundary value and material type problems.

\renewcommand{\figurename}{Figure A\hspace{-0.075 in}}
\setcounter{figure}{0}
\begin{figure*}[h!]

    \begin{center}
		\includegraphics[width=0.75\linewidth]{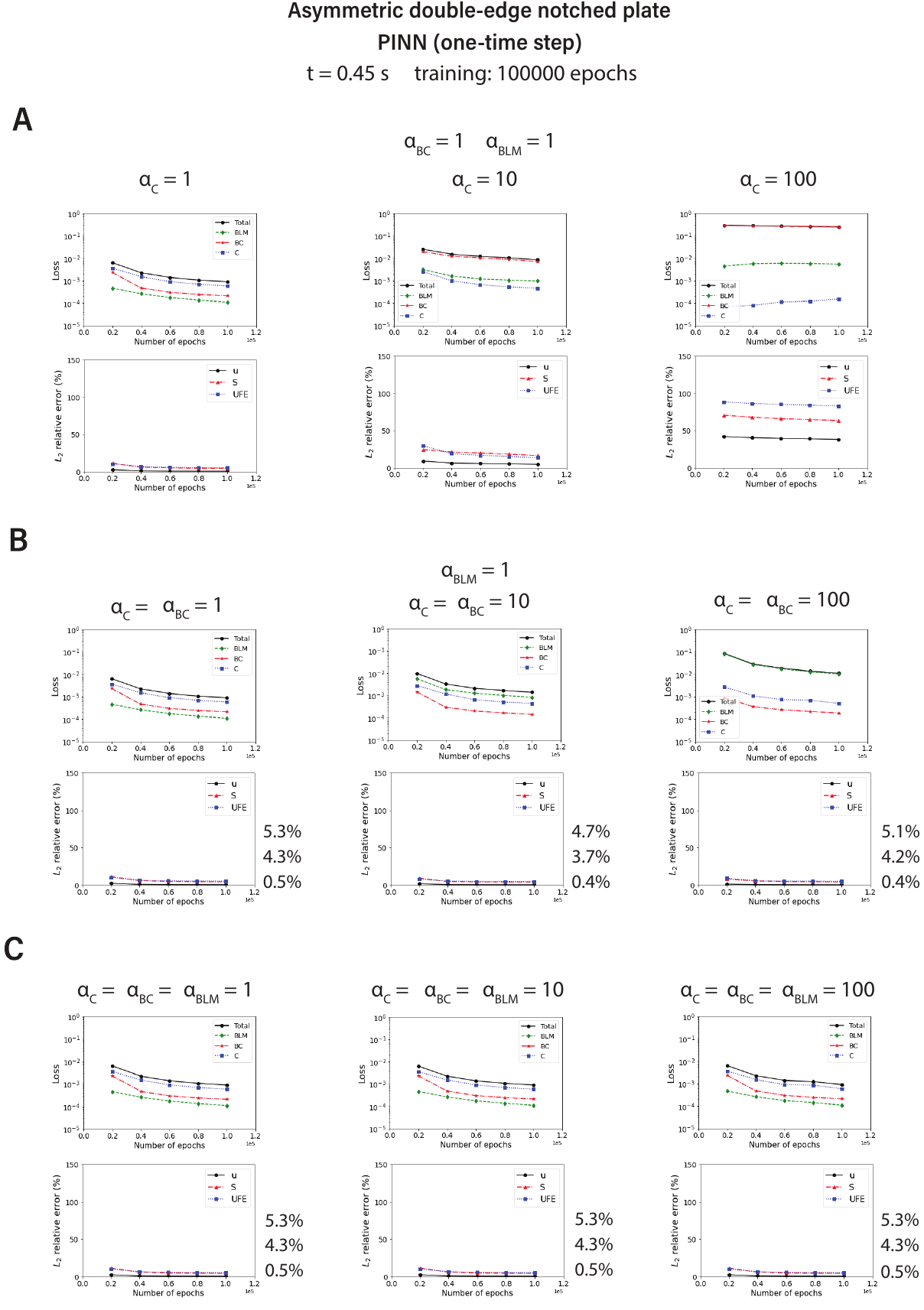}
	\end{center}
    \caption{{{\textbf{Asymmetric double-edge notched plate: effect of loss weights.} For PINN (one-time step) application, the solution is evaluated at $t=0.45$ s with 100,000 epochs training. For the baseline weight values $\alpha_{C}=\alpha_{BC}=\alpha_{BLM}=1$, $\mathcal{L}_{C}$ is significantly higher than $\mathcal{L}_{BC}$ and $\mathcal{L}_{BLM}$. The loss and $L_{2}$ error evolution with training epochs for \textbf{(A)} increase in only $\alpha_{C}$, \textbf{(B)} simultaneous increase in $\alpha_{C}$ and $\alpha_{BC}$, and (\textbf{C}) simultaneous increase in all three weights. A common observation is that the loss corresponding to the unchanged weight, and the total loss increases upon an increase in the weights. The baseline weight values provide a good balance between loss and $L_{2}$ error minimization for the material parameters used in this work.}}}
    \label{lossweighteffect}
\end{figure*}

\subsection*{PINN solution convergence}

\begin{figure*}[h!]
    \begin{center}
		\includegraphics[width=0.8\linewidth]{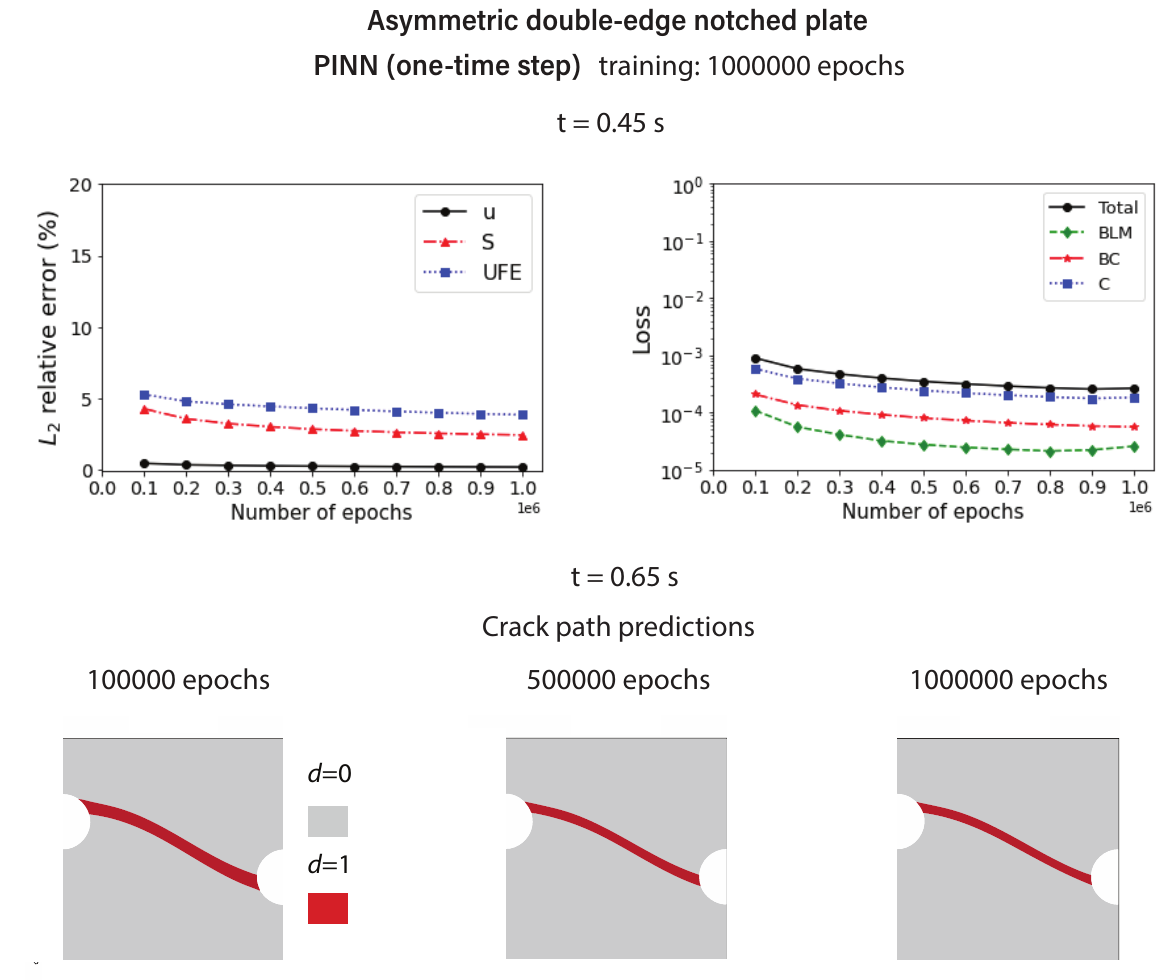}
	\end{center}
    \caption{{{\textbf{Asymmetric double-edge notched plate: PINN solution convergence.} For PINN (one-time step) application, the solution is evaluated at $t=0.45$ s and $t=0.65$ s with 1,000,000 training epochs (10 times the number of epochs used). The $L_{2}$ errors and losses decrease monotonically and approximately converge with respect to the number of training epochs. The overall width of the crack path decreases with epochs, and the path converges.}}}
    \label{PINNconvergence}
\end{figure*}

The convergence of the PINN solutions for the asymmetric double-edge notched plate geometry is verified with respect to the number of training epochs. The $L_{2}$ error, loss ($t=0.45$ s), and final crack path ($t=0.65$ s) evolution with training epochs (1,000,000 epochs - 10 times the number of epochs used) is shown in Figure A\ref{PINNconvergence}. The errors and losses decrease monotonically and approximately converge. The overall width of the crack path decreases with epochs, and the path converges. The PINN formulation does not have an inherent length scale parameter to control the damage zone width. Consequently, the predictions are a function of the neural network parameters (see Figure A5).  There is no physical correlation. Even in the baseline FEM with gradient damage approach, where a diffusive crack is considered, the ``length scale" parameter is present purely for numerical reasons.

\renewcommand{\figurename}{Figure A\hspace{-0.065 in}}

\subsection*{Comparison of PINN (incremental) and PINN (one-time step)}

\begin{figure*}[h!]
    \begin{center}
		\includegraphics[width=0.9\textwidth]{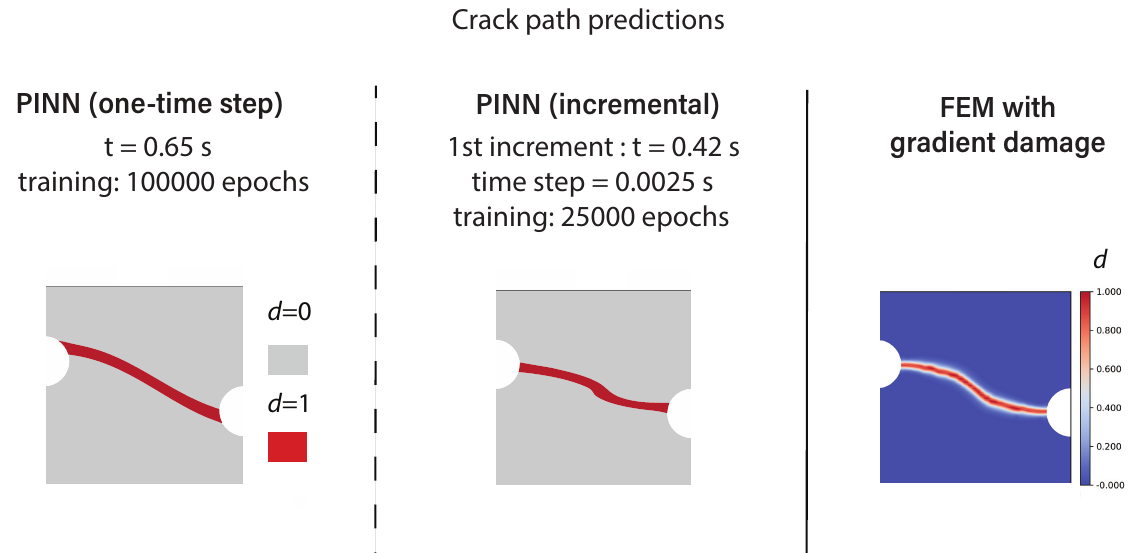}
	\end{center}
    \caption{{\textbf{Asymmetric double-edge notched plate: comparison of PINN (incremental) and PINN (one-time step) crack path predictions.} Comparison of PINN (incremental) [1st increment at $t$=0.45s, time step=0.0025 s, 25,000 epochs training] and PINN (one-time step) [$t$=0.65 s, 100,000 training epochs] crack predictions. The PINN (incremental) crack path solution shows better agreement with the baseline FEM with gradient damage solution. The computationally inexpensive PINN (one-time step), however, predicts the crack path with reasonable accuracy. PINN (one-time step) can thus enable computationally cheaper neural network parameter and architecture optimization before PINN (incremental) applications for accurate final results.}} 
    \label{two asymm holes incremental one step comp}
\end{figure*}

The crack path predictions using PINN (incremental) [1st increment at $t$=0.45 s, time step=0.0025 s, 50,000 epochs training] and PINN (one-time step) [$t$=0.65 s, 100,000 training epochs] are shown in Figure A\ref{two asymm holes incremental one step comp}. The PINN (incremental) crack path solution exhibits better agreement with the baseline FEM with gradient solution. However, the computationally inexpensive PINN (one-time step) predicts the crack path with reasonable accuracy. Hence, neural network parameter and architecture optimization can be performed in a computationally efficient manner using PINN (one-time step) before PINN (incremental) applications for accurate final results.

\renewcommand{\figurename}{Figure A\hspace{-0.06 in}}

\subsection*{Numerical factorial experiment: complete results}

\label{complete factorial results appendix}

\begin{figure*}[h!]
    \begin{center}
		\includegraphics[width=\textwidth]{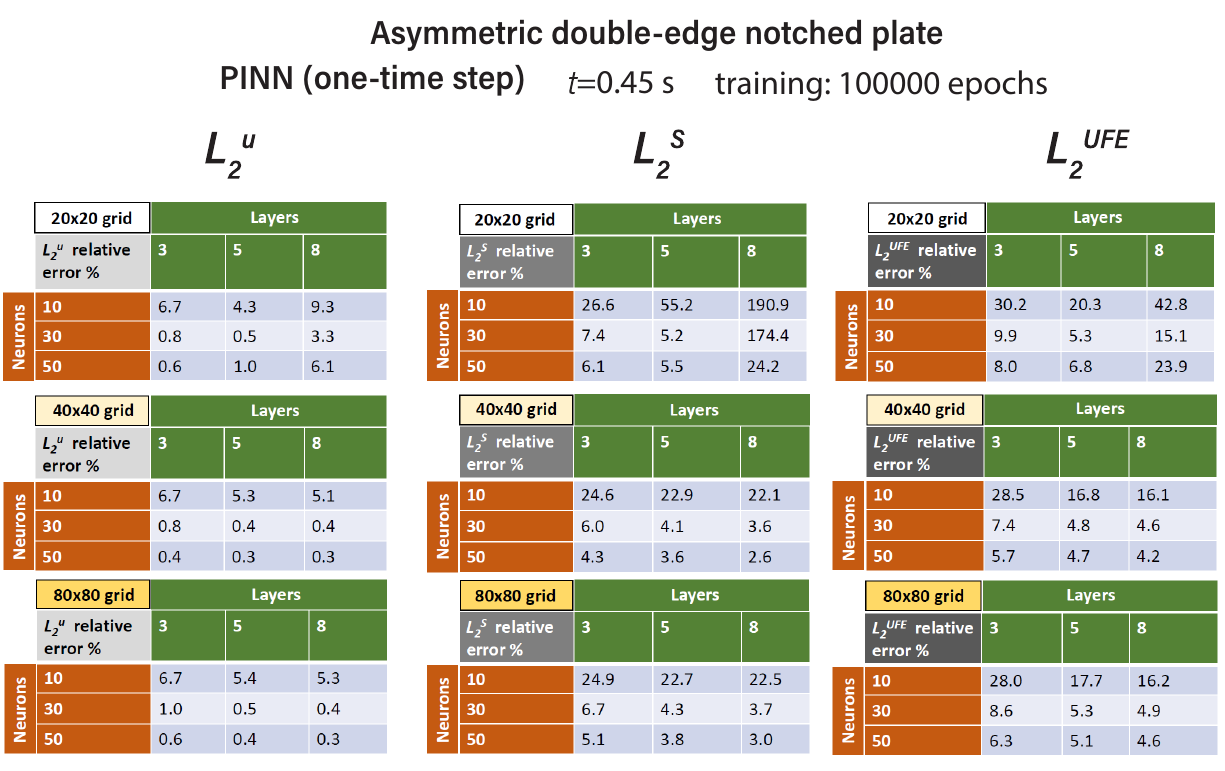}
	\end{center}
    \caption{{\textbf{Asymmetric double-edge notched plate: numerical factorial experiment ($L_{2}$ errors, $t$=0.45 s).} For $L_{2}$ error comparison, PINN (one-time step) is applied at $t=0.45$ with 100,000 epochs training. $L_2^{u}$, $L_2^{S}$, and $L_2^{\text{UFE}}$ relative errors (\%) with variations in the number of layers, number of neurons in each layer, and grid density are shown. The trends of the $L_{2}$ errors with respect to each hyperparameter considered are non-monotonic. Non-optimal hyperparameter values result in underfitting or overfitting.}} 
    \label{two asymm holes PDM-PINN (one time step) factorial bf}
\end{figure*}

\begin{figure*}[h!]
    \begin{center}
		\includegraphics[width=0.61\textwidth]{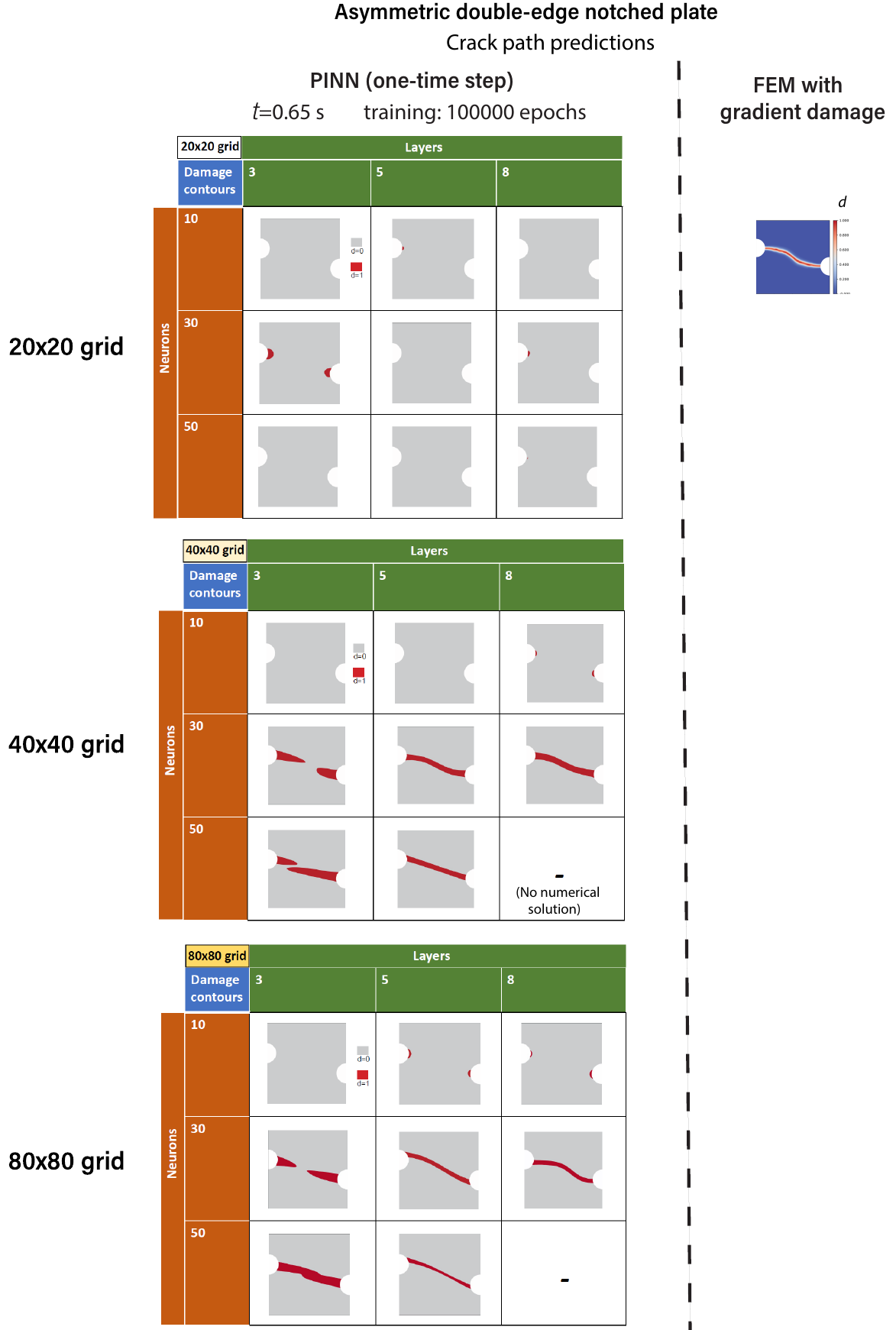}
	\end{center}
    \caption{{\textbf{Asymmetric double-edge notched plate: numerical factorial experiment (Crack path, $t$=0.65 s).} PINN (one-time step) is applied at $t=0.65$ with 100,000 epochs training for the crack path comparison. The crack paths for $20 \times 20$, $40 \times 40$, and $80 \times 80$ grid densities with variations in the number of layers and the number of neurons in each layer are shown. The evolution of the crack path with respect to each hyperparameter considered is non-monotonic. Non-optimal hyperparameter values result in underfitting or overfitting.}} 
    \label{two asymm holes PDM-PINN (one time step) factorial af}
\end{figure*}

The complete results of the numerical factorial experiment (described in Section \ref{factorial experiment}) for $L_{2}$ errors and crack paths are shown in Figures A\ref{two asymm holes PDM-PINN (one time step) factorial bf} and A\ref{two asymm holes PDM-PINN (one time step) factorial af}, respectively.  

\vspace{1 cm}

\renewcommand{\figurename}{Figure A\hspace{-0.05 in}}

\renewcommand{\figurename}{Figure A\hspace{-0.075 in}}

\hspace{-0.55 cm}\underline{Single-edge notched plate with a hole and plate with four randomly distributed holes}

\subsection*{$L_{2}$ relative error and loss evolution during training}

\label{rest two geometries errors and losses appendix}

\begin{figure*}[h!]
    \begin{center}
		\includegraphics[width=0.9\textwidth]{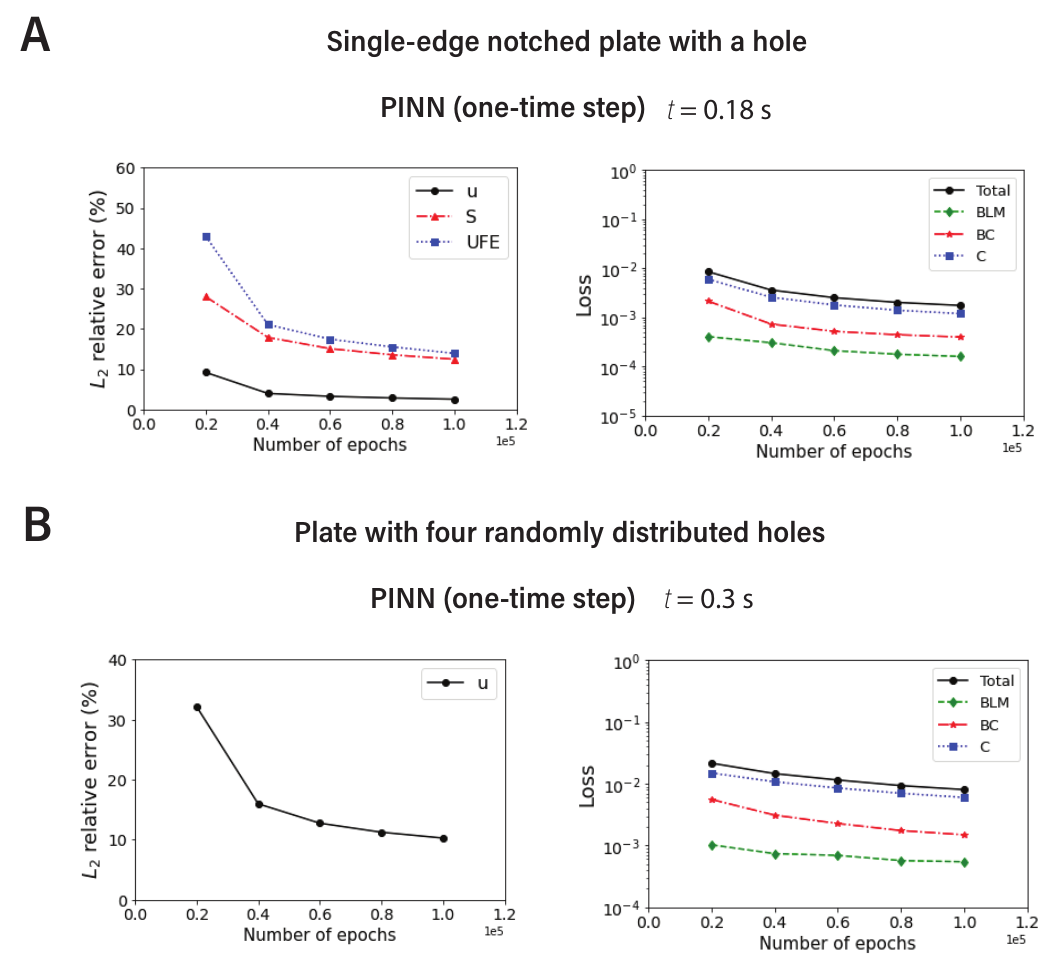}
	\end{center}
    \caption{{\textbf{Single-edge notched plate with a hole and plate with four randomly distributed holes: $L_{2}$ relative error and loss evolution during training.} PINN (one-time step) is applied at $t=0.18$ s and $t=0.3$ s (both before damage initiation) with 100,000 epochs training, respectively, for \textbf{(A)} the single-edge notched plate with a hole and \textbf{(B)} the plate with four randomly distributed holes geometries. Only $L^{u}_{2}$ is presented for the plate with four randomly distributed holes due to numerical issues in the interpolation of the relatively complex stress and undamaged free energy density fields on the testing grid. The errors and losses decrease monotonically with the number of epochs in this range for both geometries.}} 
    \label{rest geometries error and loss}
\end{figure*}

The $L_{2}$ relative error and loss evolution with the number of epochs during training for both geometries is shown in Figure A\ref{rest geometries error and loss}. For the plate with four randomly distributed holes, only $L^{u}_{2}$ is presented due to numerical issues in the interpolation of the relatively complex stress and undamaged free energy density fields on the testing grid. The errors and losses decrease monotonically with the number of epochs in this range for both geometries.

\renewcommand{\figurename}{Figure A\hspace{-0.07 in}}

\subsection*{Effect of collocation point distribution}

\label{rest two geometries structure dependence appendix}

\begin{figure*}[h!]
    \begin{center}
		\includegraphics[width=0.7\textwidth]{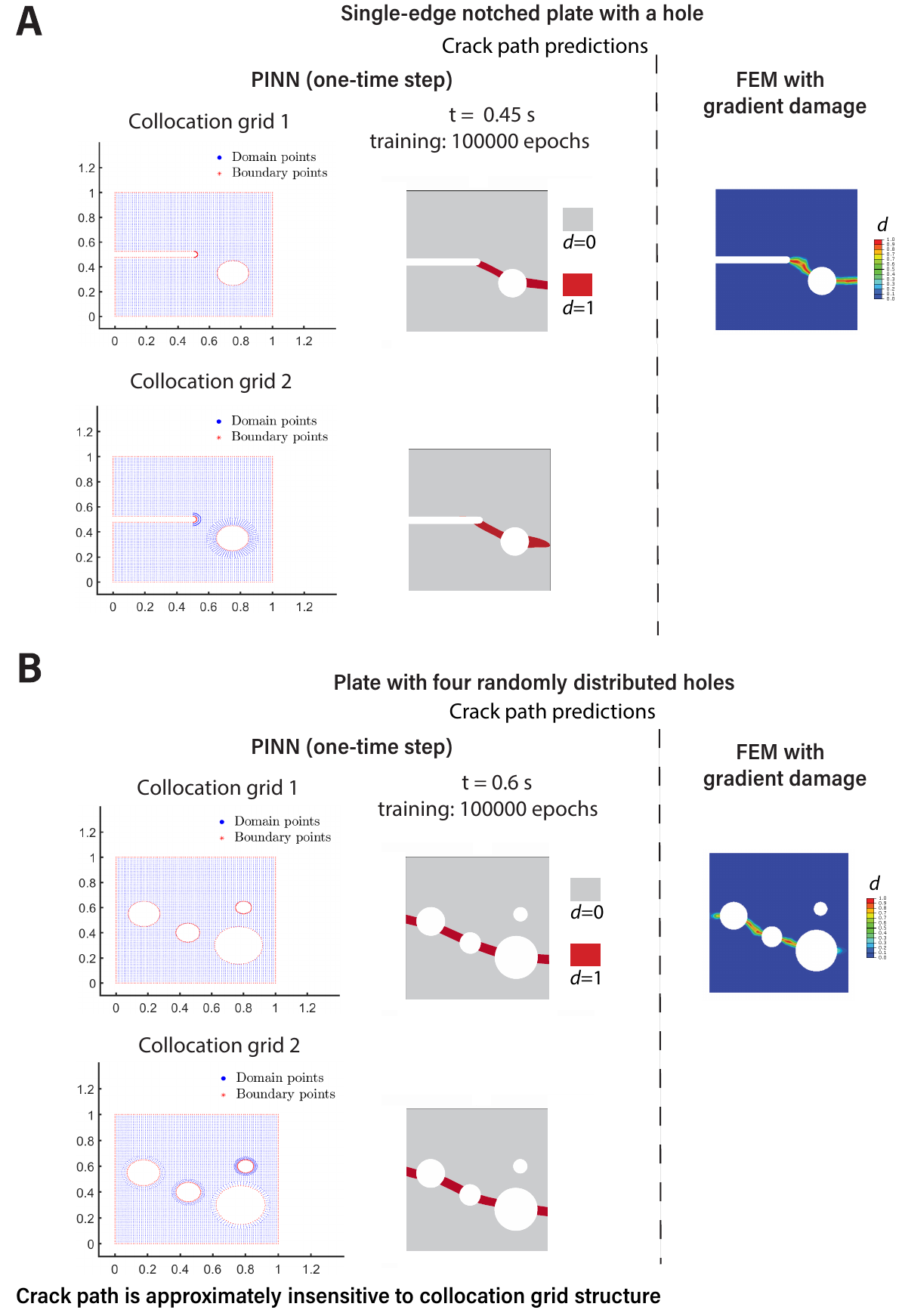}
	\end{center}
    \caption{{\textbf{Single-edge notched plate with a hole and plate with four randomly distributed holes: effect of collocation point distribution on crack path.} PINN (one-time step) is applied at $t=0.45$ s and $t=0.6$ s with 100,000 epochs training, respectively, for \textbf{(A)} the single-edge notched plate with a hole and \textbf{(B)} the plate with four randomly distributed holes geometries. The two different collocation point distributions considered for both geometries are shown. Good agreement of the PINN (one-time step) predictions with the baseline FEM with gradient damage solutions for both geometries. The PINN solution's insensitivity to the collocation point distribution is reemphasized.}} 
    \label{rest geometries grid dependence figure}
\end{figure*}

Figure A\ref{rest geometries grid dependence figure} shows the different collocation point distributions considered and the corresponding PINN (one-time step) crack path solutions for both geometries. The PINN solutions are approximately insensitive to the collocation point distribution and agree well with the baseline FEM with gradient damage solutions for both geometries.

\renewcommand{\figurename}{Figure A\hspace{-0.07 in}}

\end{appendices}

\clearpage

\bibliography{SPF_PINN}

\newpage

\end{document}